\documentclass[usenatbib]{mn2e}
\usepackage{url}
\usepackage[colorlinks=true,citecolor=blue]{hyperref}
\usepackage{graphicx,graphics,color}
\usepackage[pagewise]{lineno}
\usepackage[T1]{fontenc}
\usepackage{times}
\usepackage{ae,aecompl}
\newcommand\chandra{{\it Chandra}}

\newcommand\kms{\ifmmode {\rm~km\ s}^{-1} \else ~km s$^{-1}$\fi}
\newcommand\Hunit{\ifmmode {\rm~km\ s}^{-1}\ {\rm Mpc}^{-1}
        \else ~km s$^{-1}$ Mpc$^{-1}$\fi}
\newcommand\ctssec{\ifmmode {\rm~count\ s}^{-1} \else ~count s$^{-1}$\fi}
\newcommand\ergsec{\ifmmode {\rm~erg\ s}^{-1} \else
        ~erg s$^{-1}$\fi}
\newcommand\funit{\ifmmode {\rm~erg\ s}^{-1}\;{\rm cm}^{-2} \else
        ~ergs s$^{-1}$ cm$^{-2}$\fi}
\newcommand\phflux{\ifmmode {\rm~photon\ s}^{-1}\;{\rm cm}^{-2}
        \else   ~photon s$^{-1}$ cm$^{-2}$\fi}
\newcommand\efluxA{\ifmmode {\rm~erg\ s}^{-1}\;{\rm cm}^{-2}\;{\rm
        \AA}^{-1} \else ~erg s$^{-1}$ cm$^{-2}$ \AA$^{-1}$\fi}
\newcommand\efluxHz{\ifmmode {\rm~erg\ s}^{-1}\;{\rm cm}^{-2}\;{\rm
        Hz}^{-1} \else ~erg s$^{-1}$ cm$^{-2}$ Hz$^{-1}$\fi}
\newcommand\cc{\ifmmode {\rm~cm}^{-3} \else cm$^{-3}$\fi}
\newcommand\FWHM{\ifmmode {\rm~FWHM} \else ${\rm~FWHM}$\fi}
\newcommand\Zsun{\ifmmode Z_{\odot} \else $Z_{\odot}$\fi}
\newcommand\Lsun{\ifmmode L_{\odot} \else $L_{\odot}$\fi}

\newcommand\hbeta{\ifmmode {\rm H}\beta \else H$\beta$\fi}
\newcommand\Kalpha{\ifmmode {\rm K}\alpha \else K$\alpha$\fi}
\newcommand\nh{\ifmmode N_{\rm H} \else N$_{\rm H}$\fi}

\newcommand{\mnras}{MNRAS}

\newcommand{\mac}{\rm~MACS0417}
\newcommand{\colred}[1]{\textcolor{red}{#1}}

\title[MACS~J0417.5-1154]{A combined X-ray, optical and radio view of the merging galaxy cluster MACS~J0417.5-1154}
\author[Pandge et. al.]{M. B. Pandge,$^{1}${\thanks{Email: mbpandge@gmail.com}}, R.~Monteiro-Oliveira,$^{2,3}$ J.~Bagchi,$^{4}$ A.~Simionescu$^{5}$, \newauthor  M.~Limousin$^{6,7}$ and S.~ Raychaudhury$^{4,8,9}$\\
$^{1}$Dayanand Science College, Barshi Road, Latur, Maharashtra 413512, India\\
$^{2}$Universidade Federal do Rio Grande do Sul,  Instituto de F\'isica, Departamento de Astronomia, 91501-970 Porto Alegre, Brazil\\
$^{3}$Universidade de S\~ao Paulo, Inst. de Astronomia, Geof\'isica e Ci\^encias Atmosf\'ericas, Depto. de Astronomia, 05508-090 S\~ao Paulo, Brazil\\
$^{4}$Inter-University Centre for Astronomy and Astrophysics, Post Bag 4, Ganeshkhind, Pune 411007, India.\\
$^{5}$Institute of Space and Astronautical Science (ISAS), JAXA, 3-1-1 Yoshinodai, Chuo-ku, Sagamihara, Kanagawa 252-5210, Japan\\
$^{6}$Aix Marseille Univ, CNRS, LAM, Laboratoire  d$^{'}$Astrophysique de Marseille, Marseille, France\\
$^{7}$Aix Marseille Univ, CNRS, CNES, LAM, Marseille, France\\
$^{8}$Department of Physics, Presidency University, 86/1 College Street, Kolkata 700073, India\\
$^{9}$School of Physics and Astronomy, University of Birmingham, Birmingham B15~2TT, UK
}

\begin{document}
\pagerange{\pageref{firstpage}--\pageref{lastpage}} \pubyear{2016}
\maketitle
\begin{abstract}
  We present a comprehensive multi-wavelength analysis of the merging galaxy cluster 
  MACS~J0417.5-1154 at a redshift of $z=0.44$, using available images
 red obtained with \chandra\ in X-ray, {\it Subaru}, {\it Hubble Space
    Telescope} (HST) in optical, {\it Giant Metrewave Radio Telescope}
  (GMRT) in radio and {\it Bolocam} at 2.1~mm wavelength. This is an example of a complex merging galaxy cluster    also   hosting a steep-spectrum  Mpc scale radio halo.  
  The   mass distribution obtained by weak lensing reconstruction shows 
  that MACS~J0417.5-1154 belongs to the dissociative class of mergers,
  where one of its substructures has had its gas content detached
  after the pericentric passage. We find the main cluster mass 
   $M_{200}~=~11.5^{+3.0}_{-3.5} \times 10^{14}\ M_\odot$ and the
  smaller second (sub)cluster mass to be   $M_{200}~=~1.96^{+1.60}_{-0.95} \times 10^{14}\ M_\odot$, leading to   a large  total mass of $M_{200}~=~13.8^{+2.6}_{-2.8} \times
  10^{14}\ M_\odot$.  The overall structure, surface brightness
  profile, temperature and metal abundance of the intra-cluster medium (ICM) all point towards the
  presence of a  cold front  and  merger induced   gas-sloshing motion near the core. 
  We detect a surface brightness edge to the south$-$east   direction at a projected distance of $\sim$45 arcsec   ($\sim$255\,kpc) from the centre of this cluster. The X-ray spectral
  analysis across the inner and outer edge  allows us to confirm the
  detected edge as a cold front. The GMRT 235\,MHz observation shows a
  comet$-$like extended  sychrotron  radio halo  emission trailing behind the cold front. The peak  of the  Sunyaev-Zel'dovich decrement  is  found   displaced from the centre of X-ray
  emission, which is interpreted  as  consequence of  the merger dynamics.
  The optical HST imaging analysis of the cluster reveals  the complex
  morphology of the BCG, with three surrounding   ring-shaped  structures  with  bright  knots, which appear to be   images  of a  multiply-imaged  strongly lensed background
  galaxy. In addition  two  previously unknown  giant arcs are  found  which are all indications
  of strong  gravitational lensing in this  massive  system.  
\end{abstract}


\section[1]{Introduction}
In the hierarchical scenario of structure formation in the Universe,
galaxy clusters and their dark matter haloes grow by mergers of smaller subhaloes,
representing galaxy groups and subclusters, and by the accretion of dark
matter and diffuse gas from the cosmic web \citep[e.g.][]{2015Natur.528..105E}. 
A rare but important part of this process are the
major mergers of galaxy clusters, which represent some of the
most energetic events in the Universe, amounting to $E\geq10^{64}$
ergs. The indicators of this process of growth and merging include
features representing ram-pressure stripping and instabilities
\citep[e.g.,][]{1982MNRAS.198.1007N}, shocks and cold fronts
\citep[e.g.,][]{2007PhR...443....1M}, diffuse radio emission
\citep[e.g.,][]{2012A&ARv..20...54F}, and significant offsets between
the hot gas and dark matter substructures
\citep[e.g.,][]{2004ApJ...604..596C}. These merging galaxy clusters
are thus excellent places to investigate some of the highest-energy phenomena
known to us, including the physics of shock and cold
fronts seen in diffuse intra-cluster medium (ICM),
the cosmic ray acceleration in clusters, and
the self-interaction properties of dark matter \citep{2002NewA....7..249B,2004ApJ...606..819M}.


Within this context we present the galaxy cluster MACS~J0417.5-1154 (hereafter MACS0417). It is a  hot ($T_{\rm
  500}=11$~keV)\footnote{$T_{\rm 500}$ refers to the temperature  measured in a circle of radius $R_{\rm 500}$, within which the
    mean mass density is 500 times the critical density of the  Universe at the cluster redshift.} galaxy cluster at $z=0.443$
  \citep{2000A&AS..144..247C}. \mac~ has a total (0.3--5 keV)   X-ray luminosity and mass of ($29.1\pm 0.5\times10^{44}){~\rm erg
    s^{-1}}$ and ($22.1\pm3.9\times10^{14})$~M$_{\odot}$ respectively,   within $R_{\rm 500}$ \citep{2011A&A...534A.109P} and it is  one of
  the clusters included in the Reionization Lensing Cluster Survey  \citep[RELICS;][]{2017AAS...22920505C}. 

\mac~hosts a $\sim$1~Mpc scale   radio halo that was first reported by
\citet{2011JApA...32..529D}, and more recently explored in detail by
\citet{2017MNRAS.464.2752P} using observations with the {\it Giant Metrewave Radio Telescope} (GMRT) at
235~MHz, 610~MHz, and the Jansky Very Large Array (JVLA) at 1575~MHz and with \chandra\ in
X-rays. The extended radio halo of the cluster is coincident with the hot X-ray gas.  


The X-ray analysis of this cluster, based on \chandra\ observations,
has revealed a distinct comet like morphology of the diffuse emission,
with a compact core that is extremely X-ray
bright \citep{2012MNRAS.420.2120M}. A combined X-ray and
optical study of \cite{2012MNRAS.420.2120M} reported the possibility
of a shock front in this system.

A close look at the   spatial distribution of the various 
  components of a galaxy cluster can help 
  to characterise its
  dynamical status.  Whereas the distribution of the hot gas is
  traced by its X-ray emission, the dark matter component
  (which corresponds to upto 80\% of the total cluster mass) can be
  mapped from its gravitational influence, imprinted in the
  distortion in the shapes of background galaxies seen beyond the cluster,
  mapped in the form of weak lensing.  In the
  case of MACS0417, the cluster was included in the 
  {\it Weighing the Giants} sample \citep{WtGI, WtGIII}, but in spite of
  the clear signs of a disturbance, MACS0417 was treated as a single
  structure in this study, which yielded $M({\rm
    <1.5~Mpc})=18.9^{+2.6}_{-2.5}\times10^{14}$ M$\odot$, indicating
  that MACS0417 is a very massive system.

In this paper, we present a detailed multi$-$wavelength analysis of
\mac~by combining the existing \chandra\ X-ray observations with {\it
  Subaru} and Hubble Space Telescope (HST) optical archival
observations.Our weak lensing modelling, the first in the
  literature accounting for multiple components in this target, has provided a
  detailed anatomy of the merging system. This significantly enhances
  the opportunity to
  perform a comparison between the spatial distribution of
  the cluster components: member galaxies, intra-cluster medium and
  dark matter. We also detect several strong lensing
features, that have not been pointed out before, in the HST archival
images.

In section \S~\ref{sec:The Optical}, we outline the observations and data reduction strategy, which includes X-ray, radio and optical imaging.  The weak lensing analysis and its results are fully described in section \S~\ref{sec:weak.lensing}, while in \S~\ref{sec:XRayAnalysis}, we describe merging signature at other wavelengths. 
In section \S~\ref{sec:Discussion} and section \S~\ref{sec:conclusion} we discuss these results in their broader context, and outline our main conclusions.

Throughout the paper, we adopt a
${\rm \Lambda CDM}$ cosmology with $H_{\rm
  0}=70$~km~s$^{-1}$~Mpc$^{-1}$, $\Omega_{\rm m}=0.3$, and
$\Omega_{\Lambda}=0.7$, in which, $1\arcmin$ at the redshift
of \mac~corresponds to $\approx 340$~kpc. Unless specifically
mentioned otherwise, errors are quoted at the $90\%$ confidence level.


\section[2]{Observations and Data Reduction}
\label{sec:The Optical}
\subsection{{\it Subaru} observations}
For the weak lensing analysis, one requires deep images, with a large
field of view, in order to map the lensed background galaxies.
Multi-band observations with SuprimeCam centred on MACS0417 were obtained from the Subaru telescope archives\footnote{\texttt{http://smoka.nao.ac.jp/}}
(see Table~\ref{tab:Subaru.imaging}).  
The image reduction process used the usual semi-automated pipeline
{\sc Sdfred} \citep{sdfred1,sdfred2}, involving bias and overscan
subtraction, flat-fielding, atmospheric and dispersion correction, sky
subtraction, auto-guide masking and alignment (simultaneously for all
three filters). The images were then combined
and mosaicked into a final image for each 
filter by registering the sub-images with {\sc
  IRAF}. Photometric calibrations in the AB system \citep{Oke74} were performed by
comparing the images with the relevant standard star catalogue
\citep{Zacharias04}. Our final  image of 37.6 $\times$ 36.4 arcmin is
sufficiently large for our purposes, enabling us to access
regions at large projected distances from the centre of the cluster.

Object catalogues were constructed using {\sc SExtractor}
\citep{sextractor} in ``double image mode'', having as reference the
image in the $R_c$ filter.  Galaxies were identified according to two
complementary criteria: for magnitudes $R_C \geq 18.0$, galaxies were
defined as the objects with FWHM $>$ 1.15 arcsec, and for the brighter
objects ($R_C<18$), galaxies were identified as having {\sc
  SExtractor's} CLASS\_STAR $<$ 0.8. Stars were likewise selected by their
stellarity index or their FWHM.

\begin{table*}
  \caption{Subaru SuprimeCam
    imaging characteristics.}
\small
\begin{tabular}{lllll}\hline
Date               &Band   &   Images     & Total exposure (min)   & Seeing (arcsec)  \\
 \hline
12-12-2001 & $V$ & 6 &24   &   1.6  \\  
12-12-2001 & $I$  & 1 & 4 & 1.6\\
02-23-2004 & $R_C$ & 6 & 24& 1.1\\
02-24-2004 & $I$ & 6 & 18 & 1.0\\  
\hline
\end{tabular}
\label{tab:Subaru.imaging}
\end{table*}
\subsection{Chandra X-ray observations}
\label{sec:The X-ray}
\mac~was observed with \chandra\ twice, in VFAINT mode, between 28 to
29 October 2009, for a total of 80~ks. The observations were analyzed
using CIAO~v4.8, with CALDB~V4.6.5. The data reduction steps are the
same as those employed in the analysis of the \chandra\ data of
MACS0553.4-3342~\citep{2017MNRAS.472.2042P}. Soft proton flares were
removed from the data, point sources were subtracted, and the rescaled
instrumental backgrounds (to ensure that their count rates in
$10-12$~keV are the same as the count rates of the observations in the
same energy band) were subtracted from the images and spectra.
The X-ray point sources were detected using the {\tt
    CIAO} task {\tt WAVDETECT} with scales of 1, 2, 4, 8 pixels,
  detected at 3$\sigma$ above the background, with an absolute
  detection threshold of
  $10^{-6}$. The missing patches left after point source removal were
  replaced with the mean surface brightness level of their surroundings
  to image the
  diffuse X-ray emission around the cluster. The regions covered by
  the compact point sources were simply excluded from the spectral extraction
  regions. A summary of the two ObsIDs used in this paper is
presented in Table~\ref{tab:observations}.
\begin{table*}
\scriptsize
 \caption{{\it Chandra} Observation log}
  \begin{tabular}{llrrrrlrlr}
      \hline
    ObsID &Observing Mode & CCDs on & Starting Date & Total Time (ks)\\
    \hline
    11753  &VFAINT &0,1,2,3,6  & 2009-10-28 &54 \\
    12010  &VFAINT &0,1,2,3,6  & 2009-10-29 &26 \\
    \hline
    \end{tabular}	
  \label{tab:observations}
\end{table*}

\subsection{Radio data}
\mac~was also observed with the {\it Giant Metrewave Radio Telescope}
(GMRT) at 235 and 610~MHz in the dual-frequency mode
during 2010 October by \cite{2011JApA...32..529D}. These observations
had led to the discovery of a steep spectrum radio halo, but no signatures of
merger shock driven radio relics are seen in the outskirts of this
cluster, unlike other extreme merging systems
\citep{2011ApJ...736L...8B,2012ApJ...744...46K,2016ApJ...818..204V}. More
information related to the radio data and reduction can be found in
\citep{2017MNRAS.464.2752P}. For the present study we use the GMRT
235 MHz radio data.

\subsection{Hubble Space Telescope observations}
\mac~has been imaged with the Hubble Space Telescope (HST) with
various broadband filters. In this paper we have used archival data in
the F606W (closest to the $V$ band; PI: van der Linden) and the F160W
(closest to the $H$ band; PI: Dan Coe) filter images, for an effective
exposure time of 1910 and 1005\,sec, respectively. These images have been
used to determine the morphology of the brightest cluster galaxy (BCG)
and to look for strong gravitational lensing features.

\subsection{Bolocam observations}

\mac~has been also imaged with the Bolocam  2.1~mm wavelength (140 GHz)  
  Galaxy Cluster survey (http://irsa.ipac.caltech.edu)
  \citep{2013ApJ...768..177S},  which reveals  a strong
  Sunyaev-Zel'dovich effect (SZE) decrement signal.  Bolocam comprises of a
  144-element bolometer array at the Caltech 10.4m Submillimeter
  Observatory, at  Mauna Kea, Hawaii, with a beam of
  $58^{\prime\prime}$ (FWHM) at 140 GHz and a circular field-of-view
  of diameter 8 arcmin. In the present analysis we used
  publicly-available, filtered SZE images from this survey, of which more
  can be found in section~\S~\ref{subsec:SZ} below.
\vspace{5mm}

\section{Weak gravitational lensing analysis}
\label{sec:weak.lensing}

\subsection{Basic concepts}

In the reconstruction of the distribution of matter using
gravitational lensing, the weak regime corresponds to the linear
manifestation of the lensing, which cannot be measured using
the effect on a single galaxy. Thus, this fenomena must be quantified in
a statistical manner.  Formally, the gravitational lensing field is
described by a scalar quantity, the convergence ($\kappa$), plus a
spin-2 tensor, the shear
($\mathbf{\gamma} = \gamma_1 + i\gamma_2$),
where
$\gamma_1 = |\gamma| \cos(2\theta)$ and
$\gamma_2 = |\gamma| \sin(2\theta)$  ($\theta$ is the shear direction), both being
related to the second derivatives of the projected gravitational
potential \citep[e.g.][]{mellier99, schneider05, kneib_nata11}. For
circularly symmetric systems, the shear components are described in
terms of a tangential component (in relation to the lens centre),
frequently called $\gamma_+$, and another at 45 deg with respect to
it, known as $\gamma_\times$.

Physically, the convergence $\kappa$ is directly related to the
projected surface mass density of the lens, in units of the lensing
critical density
\begin{equation}
\Sigma_{cr} = \frac{c^2}{4\pi G} \frac{D_s}{D_{ds} D_d} \, ,
\label{eq:sig_crit}
\end{equation}
with $D_s$, $D_{ds}$ and $D_d$ being the angular diameter distances to
the source (i.e. background objects with respect to the lens), between
the lens and the source, and to the lens, respectively.  The domain of
the weak lensing regime can now be defined as the projected regions in
the image where we find $\kappa \equiv \Sigma/\Sigma_{cr}\ll 1$.
Geometrically, the convergence is responsible for an isotropic
magnification of the sources.

In an unlensed sample, the  value of average source
ellipticity is expected to be zero. The lensing effect, however,
results in coherent distortions in these images in a way that the
average ellipticity will tend to the effective shear or distortion
$g$,
\begin{equation}
\langle e \rangle \simeq g \equiv \frac{\mathbf{\gamma}}{1-\kappa} \, ,
\end{equation}
where the ellipticity modulus, in terms of the semi-major axis $a$ and semi-minor
axis $b$, is defined as
\begin{equation}
e = \frac{a-b}{a+b} \, .
\end{equation}

The so-called full ellipticity (as well as the shear and the effective
shear) cannot be defined by its modulus only, because it also has an
orientation ($\theta$, i.e., the direction of the semi-major
axis). It is commonly described as a spin-2 tensor (a ``headless''
vector), whose two components can be defined as
\begin{equation}
e_1 = e \times \cos{(2\theta)}\, 
\qquad {\rm and}\qquad
e_2 = e \times \sin{(2\theta)}\, .
\end{equation}

\subsection{Identifying the galaxy populations}

\begin{figure}
\centering
\includegraphics[angle=90, width=\columnwidth]{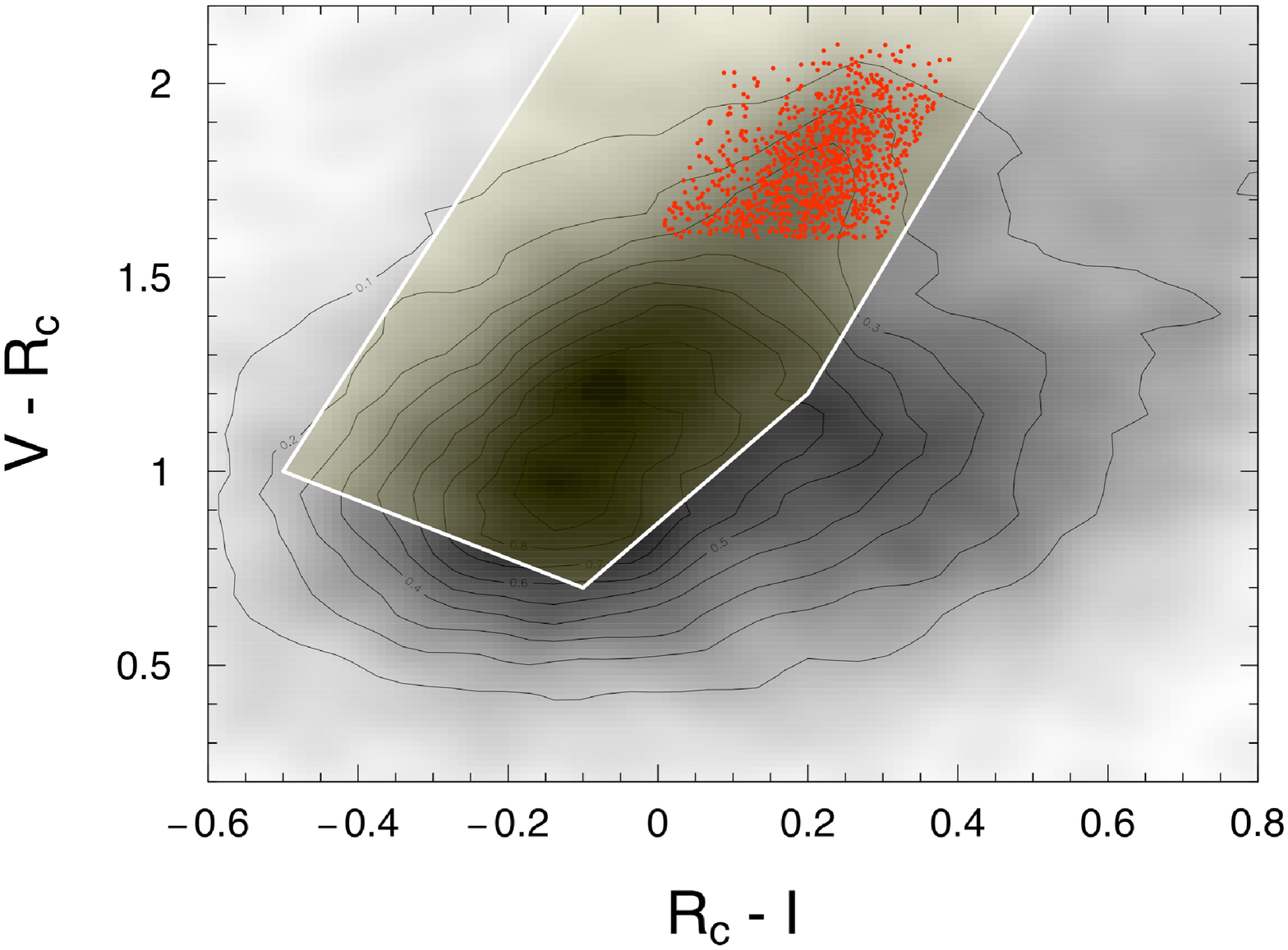}
\caption{Colour-colour diagram of the selected galaxies in the Subaru
  optical images. Black thin lines represent the density of galaxies, while
  red cluster member galaxies (red points) were found through
  a statistical subtraction process
  \citep[e.g.][]{Monteiro-Oliveira17b}. Departing from \citet{med10}
  criteria, we have identified the locus preferentially occupied by
  the foreground population (yellow central region) in the sense that
  all galaxies outside this region were considered as part of the
  background sample if $R_C>23$.}
\label{cc}
\end{figure}

The {\it locus} occupied by the red cluster members in the
colour-colour (CC) diagram \citep[e.g.][]{med10} was delineated using
a statistical subtraction process, outlined in detail in
\citet{Monteiro-Oliveira17b}. We have compared an inner region at the core
of the cluster \mac~($\sim$6 $\times$ 6 arcmin), with two outer
regions (25 $\times$ 8 arcmin each) near the image border.  We found
that the galaxy counts identified by the above process in the core
of the cluster were higher than the counts in the peripheral regions of
the Subaru image up to $R_C=23$, which was considered to be the
brightest limit for the detected background (source) galaxies. These
source galaxies are expected to occupy a region, in the CC diagram,
that is complementary to the cluster members and the foreground sample
(see Fig.~\ref{cc}).

\begin{figure*}
\centering
\includegraphics[angle=90, width=\textwidth]{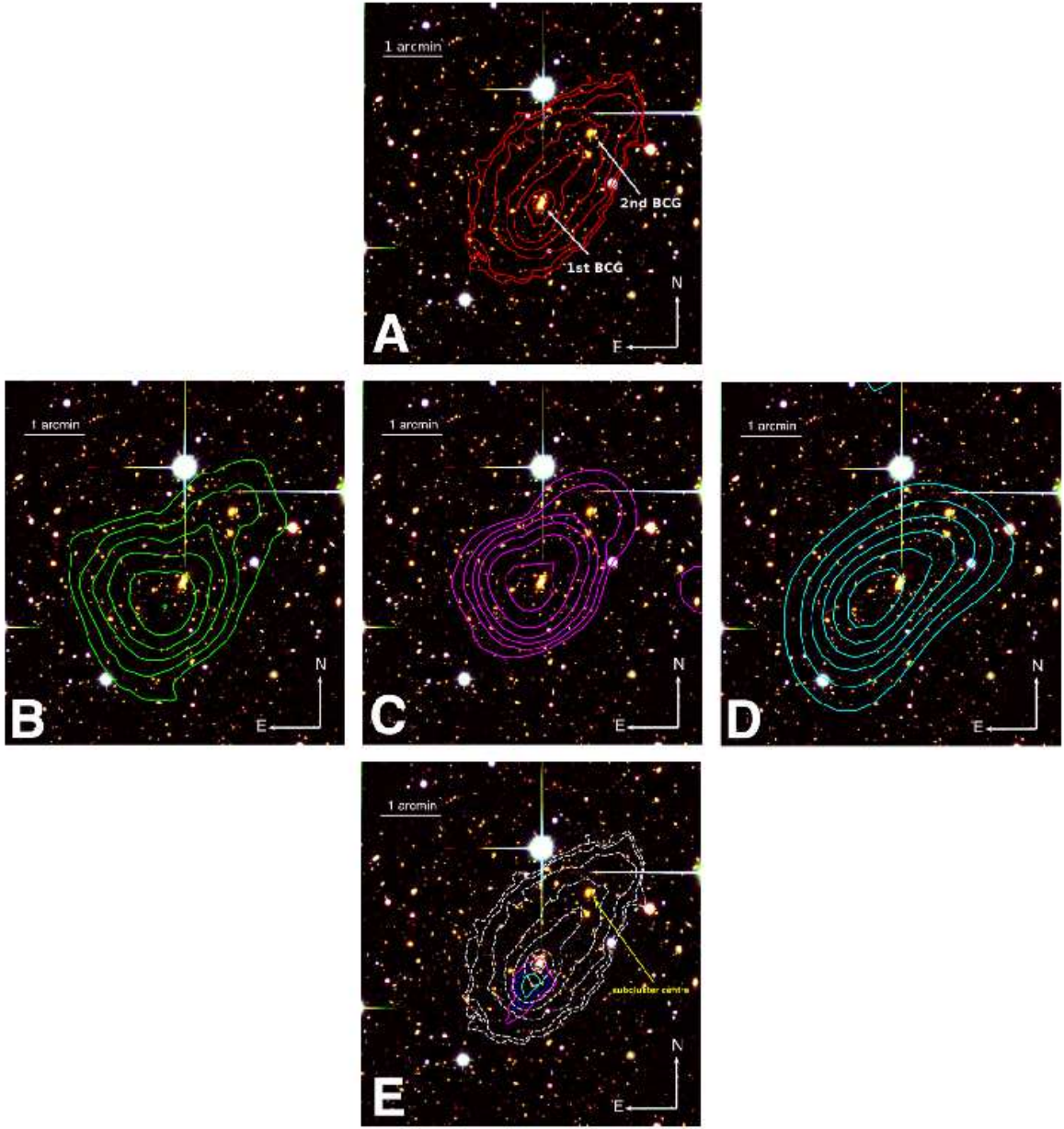}
\caption{The combined $V,R_C$ and $I$ images of the merging galaxy cluster
  \mac\ observed with SuprimeCam mounted on the Subaru
  telescope. The optical image is overlaid with (A) {\it Chandra}
  X-ray contours, (B) the logarithm of the numerical density distribution of the
  cluster red member galaxies, (C) the logarithm of the numerical density
  distribution weighted by the $R_C$ luminosity, (D) total mass
  distribution obtained by {\sc LensEnt2}, where the contours are
  linearly spaced by $0.04$ within the interval $\kappa \in
  [0.14:0.38]$ and (E) 1$\sigma$ (cyan), 2$\sigma$ (blue) and
  3$\sigma$ (magenta) confidence contours of the position of mass centre,
  and the ICM distribution traced by its X-ray emission (dotted white
  lines). The unimodal logarithmically spaced X-ray contours are nearly aligned with the axis
  connecting the two BGCs, which indicated that this might be the direction of 
  the projected merger axis. The numerical density
  distributions are very similar, and show a
  bimodal structure roughly centred on the respective BCGs. The
  recovered total mass closely follows the ICM distribution, also showing
  a unimodal peak. Finally, the modelled cluster mass centre
  (keeping the subcluster centre fixed at the 2nd BCG) shows a spatial
  agreement within 99\% confidence limit, between the 1st BCG and the
  X-ray peak, showing that this structure has not suffered a displacement of the bulk of
  the ICM.}
\label{VRI.images}
\end{figure*}
While the weak lensing technique relies only on the background galaxy
sample, we can compare further the mass distribution (dominated by the
dark matter component), with the spatial distribution of cluster
member galaxies. The resulting numerical density distribution, and the
same quantity weighted by the $R_C$ luminosity for 1014 identified red
sequence galaxies, are shown in Fig.~\ref{VRI.images} (``B'' and ``C''
respectively). In both cases, we see a bimodal distribution, closely
following the ICM distribution sampled by their X-ray emission
(``A'').  Moreover, each BCG is located near to the closest region of
higher density.  Hereafter, we will refer to the clump related to the
1st BCG (on the southeastern side)  as the ``main cluster'' and to the 2nd BCG as ``subcluster'' (on the northwestern side).

\subsection{Shape measurements}

Atmospheric blurring, combined with the angular response of telescope
optics and instrumentation, transforms the images of point-like stars
into extended images. This is quantified by the point spread function
(PSF), which is characterised with the help of the images of bright
and unsaturated stars.

The weak lensing exercise is based on shape measurements of the
background galaxies, for which the PSF
deconvolution were performed with the Bayesian code {\sc im2shape}
\citep{im2shape}\footnote{\texttt{http://www.sarahbridle.net/im2shape/}},
which models an image as a sum of Gaussians, with elliptical kernels.

\begin{figure}
\centering
\includegraphics[angle=-90, width=\columnwidth]{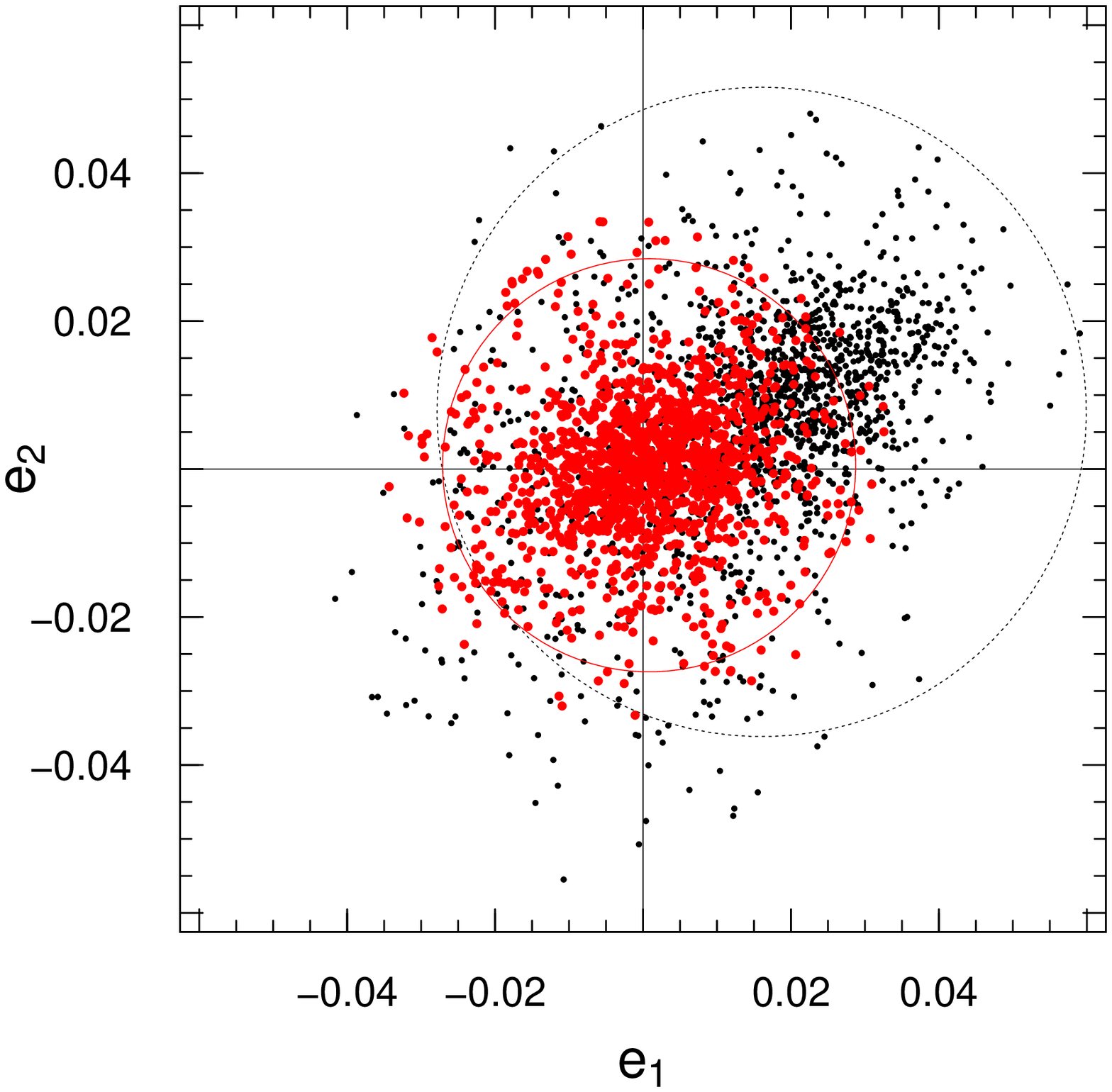}
\caption{Modelling of the PSF based on bright and unsaturated stars.
  {\it Black} points: the initial $e_1$, $e_2$ distribution with $\langle
  e_1\rangle=0.014$, $\sigma_{e_1}=0.017$ and $\langle
  e_2\rangle=0.006$, $\sigma_{e_2}=0.015$.
  If the stars were observed
  as real point sources we would expect $\langle e_1^{\star}
  \rangle=\langle e_2^{\star} \rangle=0$ so that the deviation is directly
  related to the PSF effect. {\it Red} points: the residual distribution
  after the iterative process to quantify the PSF along the field with
  $\langle {\rm res}\rangle_{e_1}=0$, $\sigma_{{\rm res}_{e_1}}=0.013$
  and $\langle {\rm res}\rangle_{e_2}=0$, $\sigma_{{\rm
      res}_{e_2}}=0.012$. The {\it circles} enclose $95\%$ of the
  points.}
\label{psf.stars}
\end{figure}

To find the PSF parameters (ellipticity components $e_1$, $e_2$ and
the FWHM), unsaturated stars are modelled as single Gaussian profiles:
for these, no deconvolution is performed. Then the discrete values are
turned continuous over the image using the thin plate spline
regression \citep[{\sc Tps},][]{fields}. Smooth PSF parameter surfaces
were reached here by setting $df=20$ parameters in the {\sc Tps}
configuration.
Aiming to remove the objects with large residuals, this process is
iterated three times, removing in each interaction 10\% of the objects
with the largest discrepancy. The final measured ellipticities, as
well the corresponding residuals, are shown in Fig.~\ref{psf.stars}.
This PSF is then used to model the shapes of the source (background)
galaxies.

We have deconvolved the local PSF to recover the distortion induced by the
gravitational lens (there is also the effect of the unknown original
shape, but it can be overcome later). As an additional precaution, we have
excluded from the final sample the galaxies with large ellipticity
errors ($\sigma_{e}>2$) and with evidence of contamination from nearby
objects. Our final background lensed galaxies consisted of $8950$
objects giving a projected density of $\sim$6 gals. arcmin$^{-2}$.

The average lensing critical surface density $\Sigma_{cr}$
(Eq.~\ref{eq:sig_crit}) was estimated by comparing our previous applied
magnitude and colour cuts (Fig.~\ref{cc}) in the COSMOS photometric
redshift catalogue \citep{cosmos}. Due to the absence of $R_C$ and $I$
bands in the survey, we have considered as proxy the filters $r'$ and $'i'$
respectively. After this procedure we found $\Sigma_{cr}=2.00(5)\times
10^{9}$ M$_\odot$ kpc$^{-2}$.

\subsection{Mass reconstruction and modelling}


We have used the code {\sc LensEnt2}
\citep{LensEnt2} designed to minimise a $\chi^2$-like statistical quantity, based
on the comparison between the measured ellipticities with those predicted by the model
 \citep[see also][]{seitz98}. As an additional input, the code
requires a smoothing scale to account for the fact 
that each galaxy is not isolated, i.e., they are correlated with those
in their vicinity. We have adopted $\sigma=80$ arcsec as used for the
numerical density maps (Fig.~\ref{VRI.images} ``B'' and ``C'').

The resulting mass distribution, presented in Fig.~\ref{VRI.images} ``D'', shows a strong similarity with the maps of
X-ray and galaxy number density. Although it does not
appear clearly bimodal, the mass distribution is elongated along the
line joining the two BCGs. This could be a hint that
\mac~represents two subclusters with widely different masses, and  that
the lensing signal of the
(much) lighter subcluster cannot be detected with the same contrast
as that of the
more massive partner.

The mass distribution in the cluster field was modelled as two lenses,
each one following a
NFW \citep{nfw96,nfw97} profile \citep{wright00}, which essentially requires four
parameters: the mass scales ($M_{200}^c$,  $M_{200}^s$) and the respective halo
concentrations ($c^c$, $c^s$). In addition, the parametrization of each halo
depends on the position of their centre ($x,y$), which can be also included as a
free parameter \citep[e.g.][]{Monteiro-Oliveira17a}. However, 
increasing the number of parameters yields output quantities which are
less  constrained. In the following, we will  implement
strategies to diminish the number of final parameters.

The halo concentration was fixed based on the mass-concentration
relationship presented by \cite{duffy08},
\begin{equation}
c=5.71\left(\frac{M_{200}}{2\times10^{12}h^{-1}M_{\odot}}\right)^{-0.084}(1+z)^{-0.47},
\label{eq:duffy_rel}
\end{equation}
where $z$ is the redshift of \mac. As suggested by
Fig.~\ref{VRI.images} (``D''), the subcluster appears to be
overwhelmed by the main cluster, which indicates that we might not be
able to fit simultaneously  both the mass and the position of the
centroid of the subcluster. In fact, initial tests have shown that
keeping the subcluster centre as a free parameter dramatically
increases the error bar in the other parameters and returned
non-convergent results. We decided then to adopt the reasonable
assumption that the centroid of mass of the subcluster is coincident
with the 2nd BCG, as suggested by the numerical density map of the red
members (Fig.~\ref{VRI.images} ``B'' and ``C''). After these
considerations, our fiducial model was reduced to 4 unknown
parameters: $M_{200}^{c}$, $M_{200}^{s}$, $ x^c$, $ y^c$, where the
index ``c'' refers to the main cluster and ``s'' to the subcluster.

Due to the absence of circular symmetry (since we are 
simultaneously modelling two halo profiles), we worked with Cartesian components
of the effective shear rather than the usual tangential components. The
transformation is done just by multiplying $g_+$ by the lensing
convolution kernel,
\begin{equation}
D_1 = \frac{y^2 - x^2}{x^2 + y^2}\, 
\quad\mbox{and}\quad
D_2 = \frac{2xy}{x^2 + y^2} \, , 
\end{equation}
where $x$ and $y$ are the Cartesian coordinates relative to the respective lens centre. 

The total effective shear calculated at each lensed  galaxy position can be
written as a sum of the contribution of each halo,
\begin{equation}
g_i =  g_{i}^c+g_{i}^s\mbox{,}\quad i\in\{1,2\}\mbox{.}
\label{sum.g}
\end{equation}
The model prediction is then compared to the measured ellipticities
$e_1$ and $e_2$ for the $N$ background sources through the $\chi^2$,
\begin{equation}
\chi^2=\sum_{j=1}^{N} \sum_{i=1}^{2}  \frac{(g_i-e_{i,j})^ 2}{\sigma_{int}^2+\sigma_{ obs_{i,j}}^2}\mbox{,}
\label{chi2}
\end{equation}
with $\sigma_{ obs_{i,j}}$ being the errors reported by {\sc im2shape}, and
$\sigma_{int}$ the uncertainty associated with the intrinsic
ellipticity distribution of the source galaxies, estimated as 0.35 for
our dataset.

Finally, the posterior can be written as 
\begin{equation}
{\rm Pr}(\rm M| data) \propto  \mathcal{L}({\rm data}|M)\times\mathcal{P}(M)\mbox{.}
\label{eq:posterior}
\end{equation} 
A uniform prior $\mathcal{P}(M):0<M_{200}\leq 9\times 10^{15}$
$M_{\odot}$ was adopted to allow the chains to quickly reach the
stationary state. For data normally distributed around the model, the
likelihood is related to the previous $\chi^2$ statistics as
\begin{equation}
\ln \mathcal{L} \propto - \frac{\chi^2}{2}\mbox{.} 
\label{likelihood.d}
\end{equation}

\subsection{Results}
\label{sec:WLR}

The posterior of the problem (Eq.~\ref{eq:posterior}) was evaluated by
an MCMC (Markov chain Monte Carlo) algorithm with a simple Metropolis
sampler implemented in the {\sc R} function {\sc MCMCmetrop1R}
\citep{MCMCpack}. We generated four chains of $1 \times 10^5$ points,
added with the first $1\times 10^4$ points as ``burn-in'' iterations,
to ensure that the chains fully represent the stationary state. To check
the final combined chain convergence, we also measured the potential
scale factor $R$, as implemented in the {\sc Coda} package which has
shown that the convergence cannot be discarded within 68\% confidence
limit ($R \le 1.0$).

\begin{figure}
\centering
\includegraphics[angle=270, width=\columnwidth]{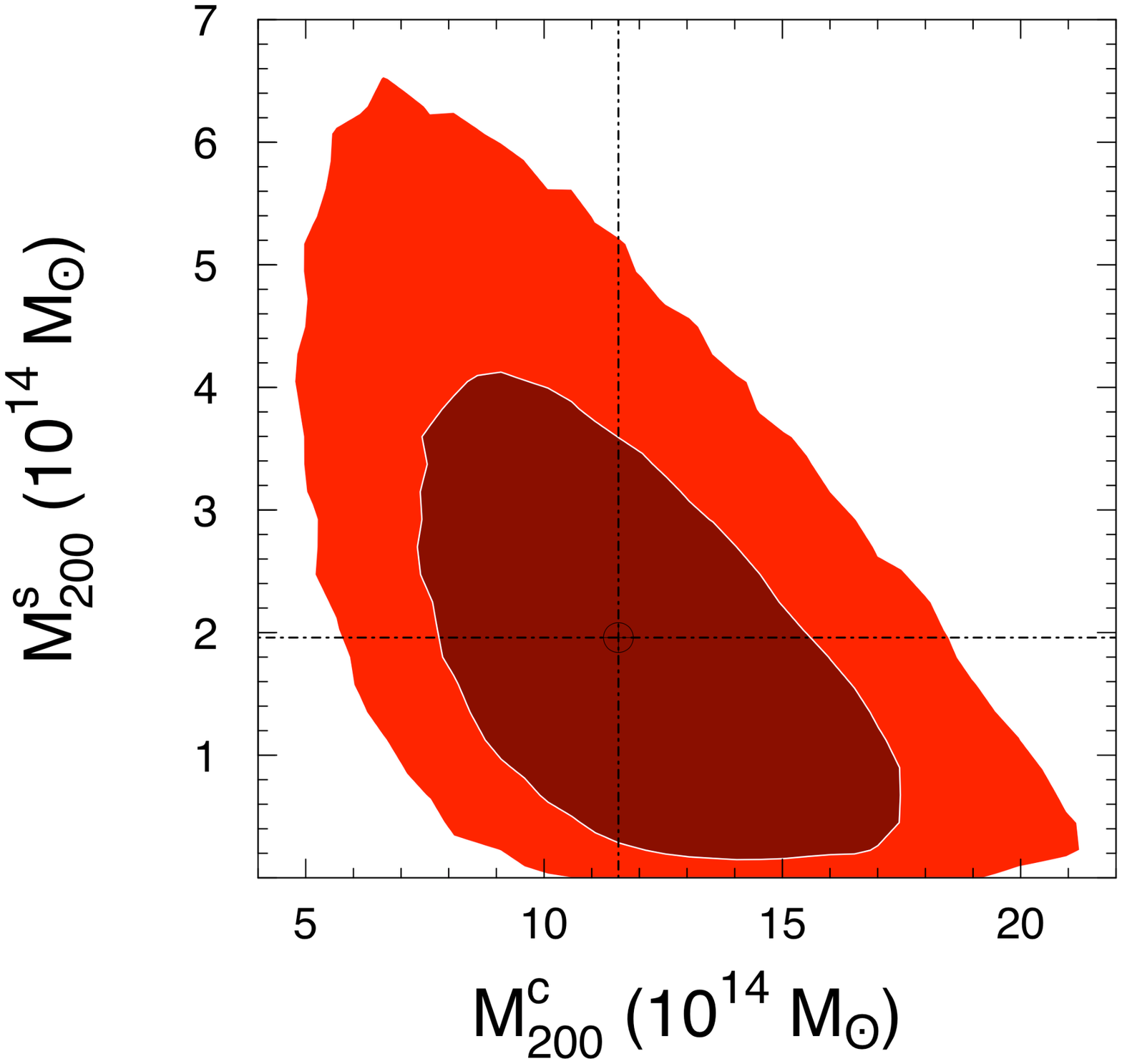}
\caption{Marginalized posterior of the main cluster (c) and
    subcluster (s) masses. The crimson area corresponds to a
    confidence level of 68\% whereas the lighter red area covers up to     the level of 95\%. The individual values (dashed-dotted lines),
    corresponding to the median of each distribution, are listed in
    Tab.~\ref{tab:masses}.  We found the Spearman correlation
    coefficient between the two parameters to be $\rho=-0.5484\pm0.0011$.}
\label{posterior.mass}
\end{figure}

The final posterior of the mass measurements is shown in
Fig.~\ref{posterior.mass}. The main cluster and subcluster masses,
marginalized over the other parameters, are presented in
Tab.~\ref{tab:masses}. Concerning these values, the previous
speculation about the high mass ratio of MACS0417 is plausible according to our model which indicates a ratio $\sim$6:1, considering the representative median values, pointing that the system has been experiencing a minor merger \citep{Martel14}. Actually, this classification is consistent with 67\% of the realizations, whereas in $\sim9\%$ the scenario is better described as a major merger (i.e. the mass ratio is less than 2:1). The corresponding total
mass\footnote{\colred{Taken as the median of the sum of both main and subcluster individual posteriors.}} is $M_{200}=13.8^{+2.6}_{-2.8} \times 10^{14}$ M$_\odot$ inside
$R_{200}=1.96\pm0.14$~Mpc. These values are found to be practically
unchanged, when we used a conservative model, keeping both the main
cluster and subcluster mass centres fixed at their corresponding
BCGs. Besides, our total mass estimation agrees within the error bars
with those obtained by \cite{WtGIII}, $M({\rm
  <1.5~Mpc})=18.9^{+2.6}_{-2.5}\times10^{14}$ M$\odot$, in spite of
the fact that their model considers MACS0417 to be a single cluster.

\begin{table}
\caption{Mass modelling results. The representative values correspond
  to the median of each marginalized PDF. The error bars refer to 68\%
  c.l.}
\begin{center}
\small
\begin{tabular}{ll}\hline
& $M_{200}$ ($10^{14}$ M$\odot$)   \\
\hline
Main cluster (c)& $11.5^{+3.0}_{-3.5}$ \\[5pt] 
Subcluster (s)   &  $1.96^{+1.60}_{-0.95}$ \\[5pt] 
Total (c+s)            &  $13.8^{+2.6}_{-2.8}$ \\
\hline
\end{tabular}
\label{tab:masses}
\end{center}

\end{table}

The confidence contours of the modelled main cluster mass centre are
presented in Fig. ~\ref{VRI.images} ``E''.  This figure also enables
us to make a comparison between the 1st BCG and the X-ray peak
positions. As seen, for example,  in the merging galaxy clusters A1758
\citep{Monteiro-Oliveira17a} and A3376 \citep{Monteiro-Oliveira17b},
we notice no significant offset among the main cluster components,
meaning that the structure has retained its gas content after the
collision with the subcluster. In fact, both locations of the
X-ray peak and the 1st BCG are consistent with the position
of the mass centre within 99 \% c.l. As far as the subcluster is
concerned, the X-ray map (Fig.~\ref{VRI.images} ``A'') shows no evidence of a related
emission peak, suggesting that it was previously disrupted due to the
large$-$scale collision \citep[as seen for example in the merging cluster A3376;][]{Monteiro-Oliveira17b}.

\section{Merging signatures at other wavelengths}
\label{sec:XRayAnalysis}
\subsection{X-ray}
\subsubsection{Imaging analysis}
\begin{figure}
\includegraphics[width=9cm,height=8cm]{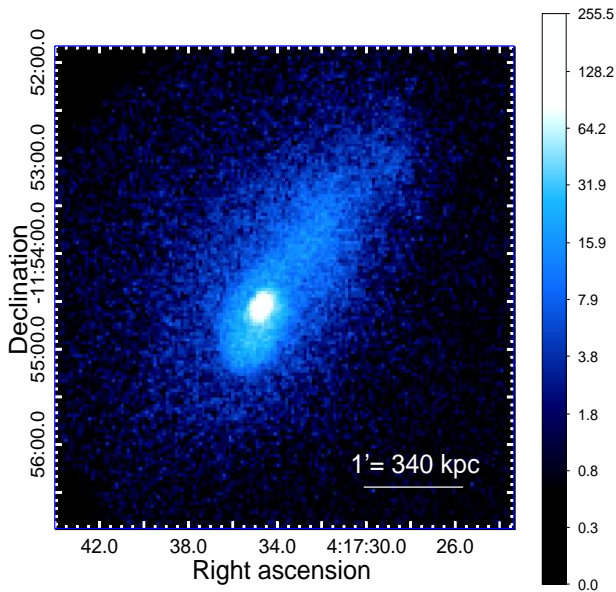}
\caption{X-ray (\chandra) log scaled surface brightness map of \mac\ in the energy band   $0.5-3.0$~keV. The image was corrected for exposure, vignetting, and after   subtracting point sources and  the instrumental background, was smoothed with a Gaussian   kernel of width 1~arcsec (2 pixels).  }
 \label{fig1} 
\end{figure}

In Fig.~\ref{fig1}, we show the X-ray surface brightness map of the
diffuse emission of the ICM of the cluster, in the energy band
0.5$-$3\,keV for the \chandra\ observation. The image has been
corrected for exposure and vignetting, with the instrumental
background subtracted, and it has been smoothed with a Gaussian kernel
of width 1~arcsec. From visual inspection of this image, it is evident that the
appearance of \mac~cluster has a comet-like morphology, with a bright
core in the X-rays. This morphology clearly indicates that a high
velocity encounter occurred in this system, similar to that in 
the Ophiuchus Cluster \citep{2010MNRAS.405.1624M} and Abell~2146 complex 
\citep{2013A&A...556A..44R}. Apart from the bright
X-ray core, the image reveals a tail-like feature directed towards the
north-east (hereafter NE) and another interesting sharp edge-like
feature evident to the south-east (hereafter SE) from the centre of
the cluster. There is a clear indication of a surface brightness
discontinuity/edge along the SE direction. To detect and confirm the
surface brightness edge, we have used various image analysis
techniques, which we discuss in detail below.

\begin{table*}
\caption{Best-fit values of the broken power-law density model}
\small
\begin{tabular}{lllllllllll}\hline
Angle               &$\alpha$1     &   $\alpha$2   & $r_{sh}$     &~~$n_0$         &C             &$\chi^{2}$/dof     \\
                    &              &               &(arcmin)      &($10^{-4} {\rm cm^{-3}}$)   &&     \\ \hline
 (180\degr$-$270\degr)&$0.46\pm0.04$ &$1.19\pm0.11$ &$0.72\pm0.03$ &$9.1\pm0.59$ &$2.21\pm0.12$  &78.81/33          \\  
\hline
\end{tabular}
\label{tab:jumps}
\end{table*}
\begin{table*}
\caption{Best fit spectral properties of the T2 and T1 regions shown in Fig.~\ref{fig:confirmed-jumps}.}
\begin{center}
\begin{tabular}{crcccccc}
\hline
\hline
Reg.    &Net    & ${\rm N_{H}}$   & ~\rm kT  & Z &  Norm ($10^{-4}$)& $P$ &\(\chi^2\) (d.o.f.) \\
        &Counts & $ {\rm10^{20} cm^{-2}}$ & \(\mbox{\rm(keV)}\)& \Zsun & ($\rm cm^{-3}$)& {($10^{-2} \rm KeV cm^{-3})$}& \\
\hline
T2 & 3160  & 7.12$\pm0.13$     & $7.20^{+1.38}_{-0.99}$& $0.54\pm0.14$  & $4.10\pm0.16$ &4.96$\pm0.92$   &62.98 (65) \\
T1 & 1350  &  8.71$\pm0.06$    & $16.37^{+6.89}_{-4.35}$& $0.97\pm0.79$	 & $1.56\pm0.20$ &4.86$\pm1.20$   &34.23 (39) \\
\hline
\hline
\label{fit_results}
\end{tabular}
\end{center}
\end{table*}
\subsubsection{Surface Brightness Profiles}
\begin{figure*}
\centering
\includegraphics[scale=0.4]{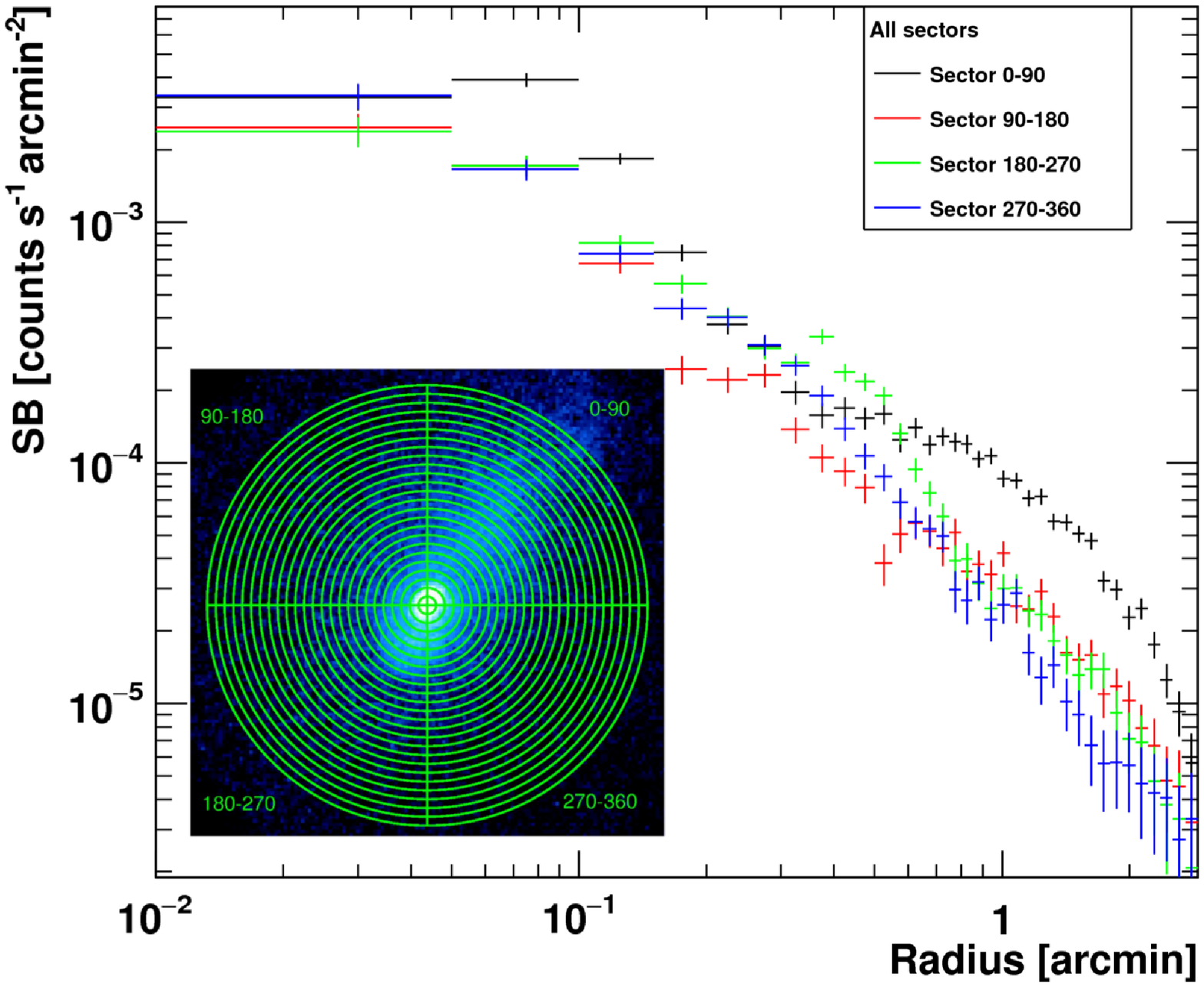}
\caption{The extracted X-ray surface brightness profiles across four sectors from the centre of \mac. The sectors and the angles (measured counter-clockwise from the Right ascension axis) used to extract the surface brightness profiles are shown in the inset.} 
\label{fig:allsec}
\end{figure*}

X-ray surface brightness profiles can reveal features like surface
brightness edges, pre- and post-shock signatures, cold fronts and
other attributes of merging events seen in galaxy clusters. We derived
surface brightness profiles from the \chandra\ observation, by
extracting X-ray counts from four sectors, from the centre of \mac~as
shown in Fig.~\ref{fig:allsec}. The sectors and 
  angles (measured counter-clockwise from the Right Ascension axis)
  used to extract the surface brightness profiles are given in the
  inset of Fig.~\ref{fig:allsec}.

\begin{figure*}
    \includegraphics[width=9.2cm,height=9.2cm]{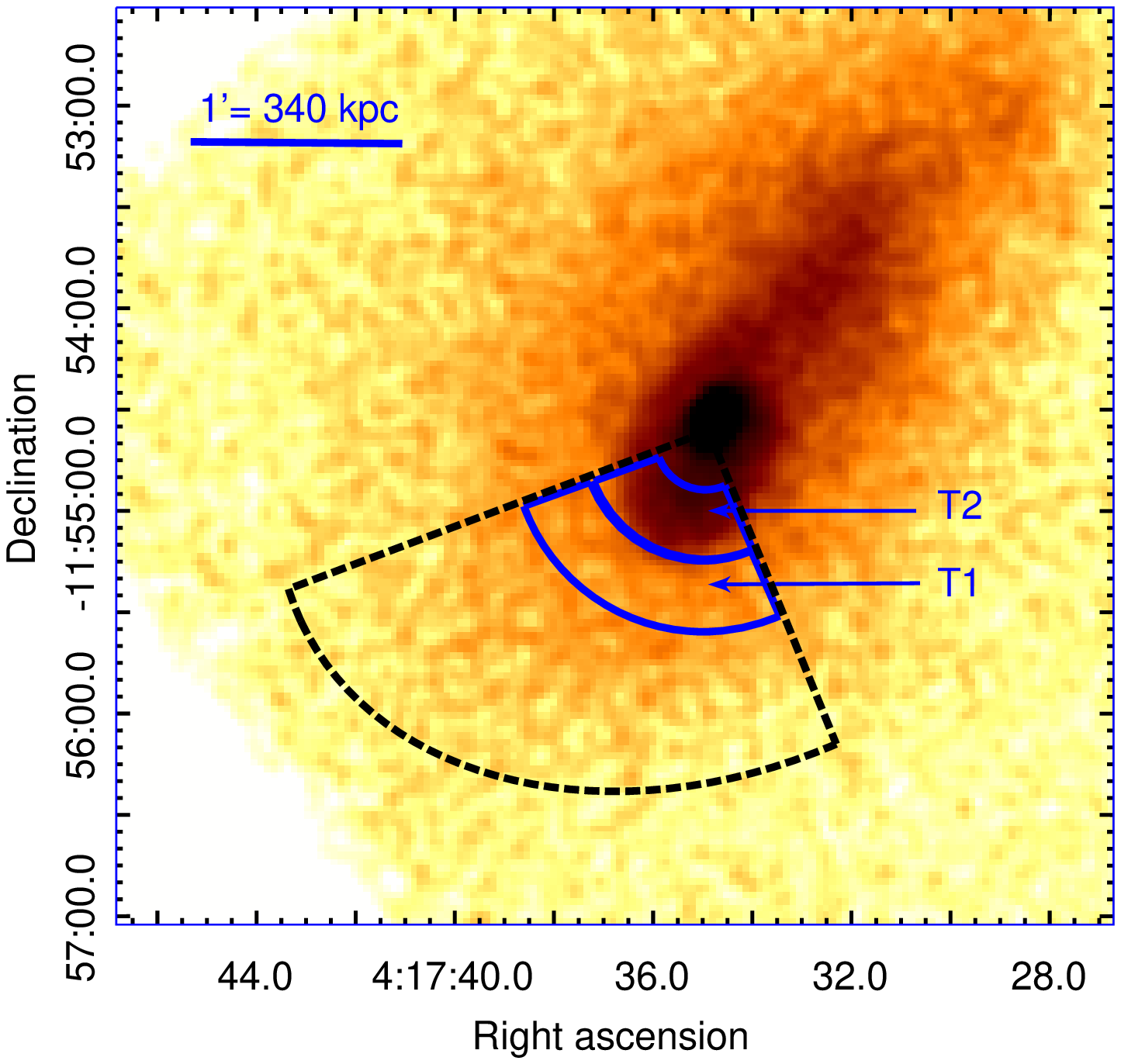}
    \hspace{-1cm}
    \includegraphics[width=8cm,height=8cm]{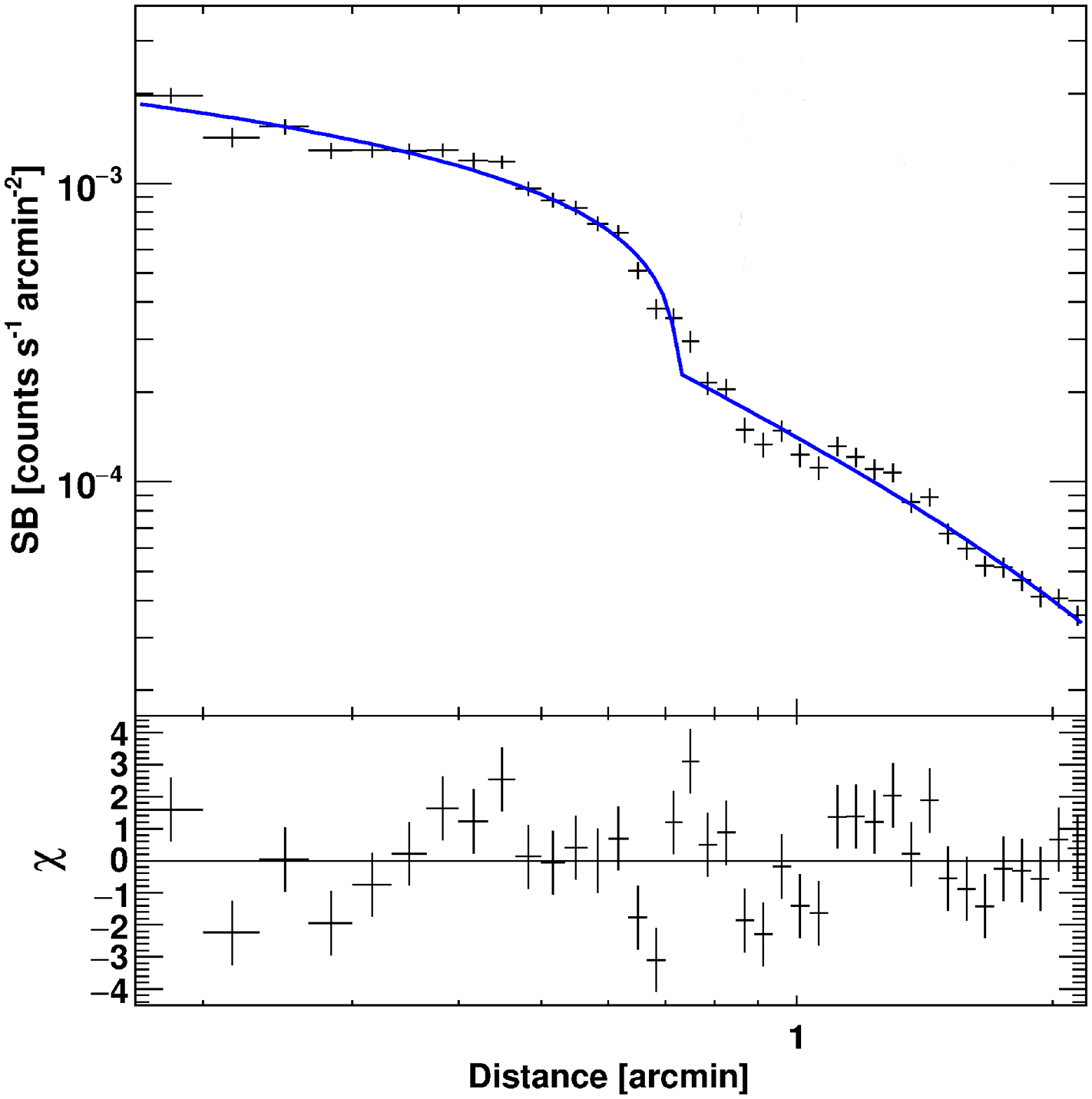}
    \caption{\emph{left panel:} The \chandra\ X-ray (0.5$-$3.0\,keV energy band) image on which the sector, used to extract the surface brightness profile
      across the SE edge, has been overlaid. The far inner
      regions, T2 and T1, were used to extract the spectrum across the
      detected surface brightness edge. \emph{Right panel:}
      Best-fit broken power-law density model (solid blue line)
      for the surface brightness edge extracted from 180\degr-270\degr from the centre of the
      cluster. The surface brightness profile is instrumental
      background subtracted. 
\label{fig:confirmed-jumps} 
\vspace{0.3cm}}
\end{figure*}

This figure clearly shows an excess X-ray emission across the position
angle (0\degr$-$90\degr) (colour: black) within the range of 20~arcsec
to 2~arcmin in radius. In the same figure, the profile extracted from
the SE direction, covering the position angle (180\degr $-$ 270\degr)
(colour: green) also shows excessive X-ray emission in the range of
radial distance between 0.1--0.5$\arcmin$ in this sector, compared
with the similar range in other profiles. Further, the surface
brightness drops abruptly at 0.5\arcmin, beyond which its behaviour is
similar to the surface brightness profiles at other ranges in azimuth. This
indicates the possible presence of a density discontinuity in this direction.

\subsubsection{A SE Surface brightness edge}
\label{sec:edges} 
We further investigated the possible density discontinuity by
extracting a surface brightness profile using a elliptical sector with 
covering the position angle (180$-$270\degr), fixing ratio between major and minor axis $\sim$ 1.13  and shown by dashed lines in Fig.~\ref{fig:confirmed-jumps} ({\it left panel}).  The surface
brightness profile across the edge, shown in the right panel, was fitted with a broken power-law density model. The broken power-law density model was parametrized as:
\begin{equation} n(r)=\left\{\begin{array}{ll} 
\mathcal{C}n_{0}~(\frac{r}{r_{sh}})^{-\alpha1}, & \mbox{ if }\hspace{2mm}r 
<r_{\rm sh}\\\\
\mathcal{}n_{0}(\frac{r}{r_{\rm sh}})^{-\alpha2} , & \mbox{if}\hspace{2mm}  r > 
r_{\rm sh}\end{array}\right.
\end{equation}
where $n$ is the electron number density, $n_{0}$ is the electron density normalization, $C$ is the density compression, $\alpha1$ and $\alpha2$ are the power-law indices,  and $r_{\rm sh}$ is the radius of the density jump. We binned the profile to have approximately $28$ counts/bin and fitted it
with the \textsc{proffit} package V1.4
\citep{2011A&A...526A..79E}. The best-fitting parameters are
summarized in Table~\ref{tab:jumps}.

This modelled density was squared and integrated along the line of
sight, assuming an isothermal plasma and a ellipsoidal geometry.

\begin{figure}
\centering
\includegraphics[width=7.5cm,height=7.5cm]{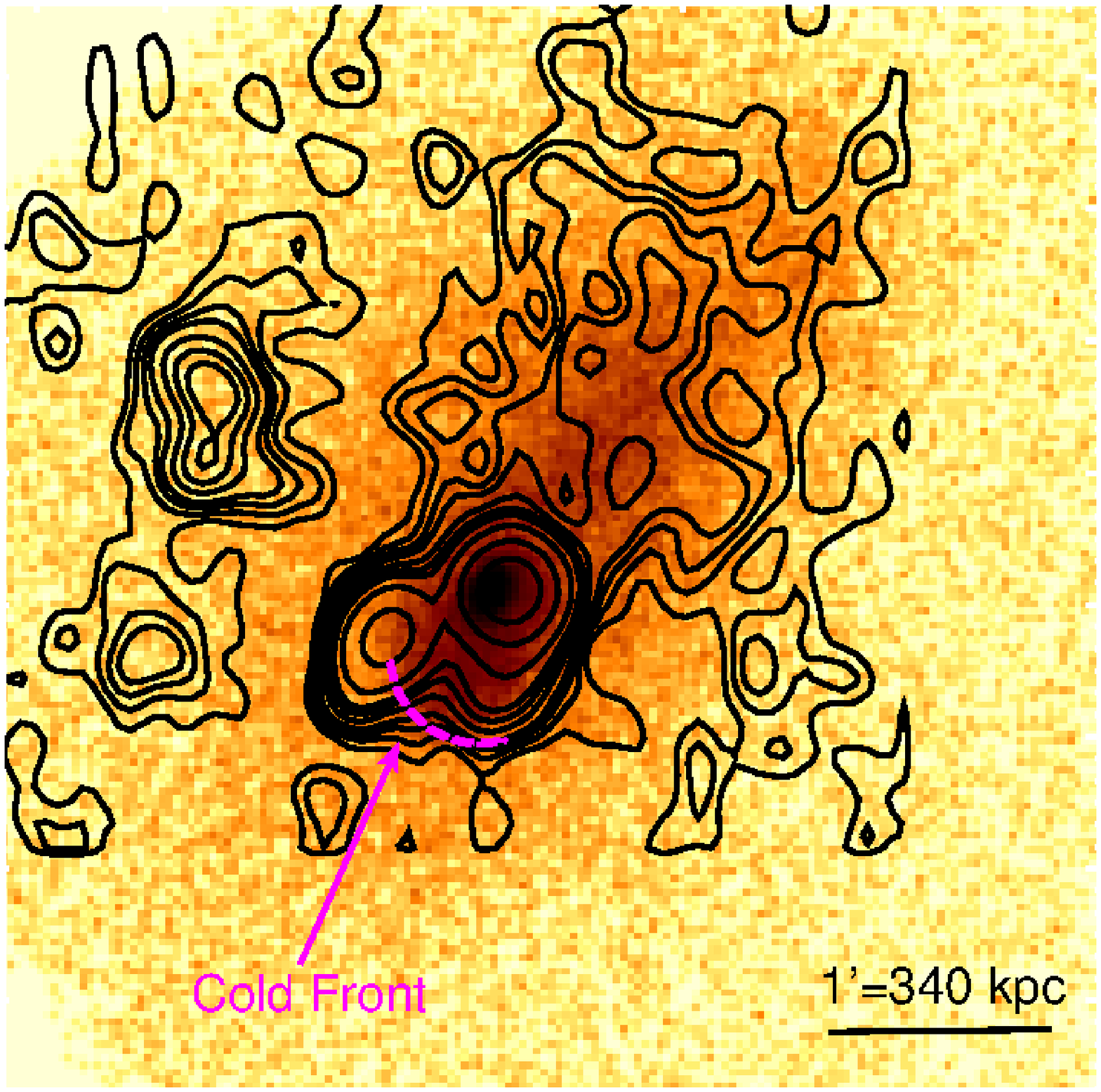}
\caption{\chandra\ 
  X-ray (0.5$-$3.0\,keV energy band) 
  surface brightness image on which the GMRT 235~MHz radio contours
  are overlaid.  The position of detected cold front (dashed magenta
  coloured arc) is shown by an arrow.}
\label{XOR}
\end{figure}
\begin{figure}
\includegraphics[width=9.5cm,height=7.6cm]{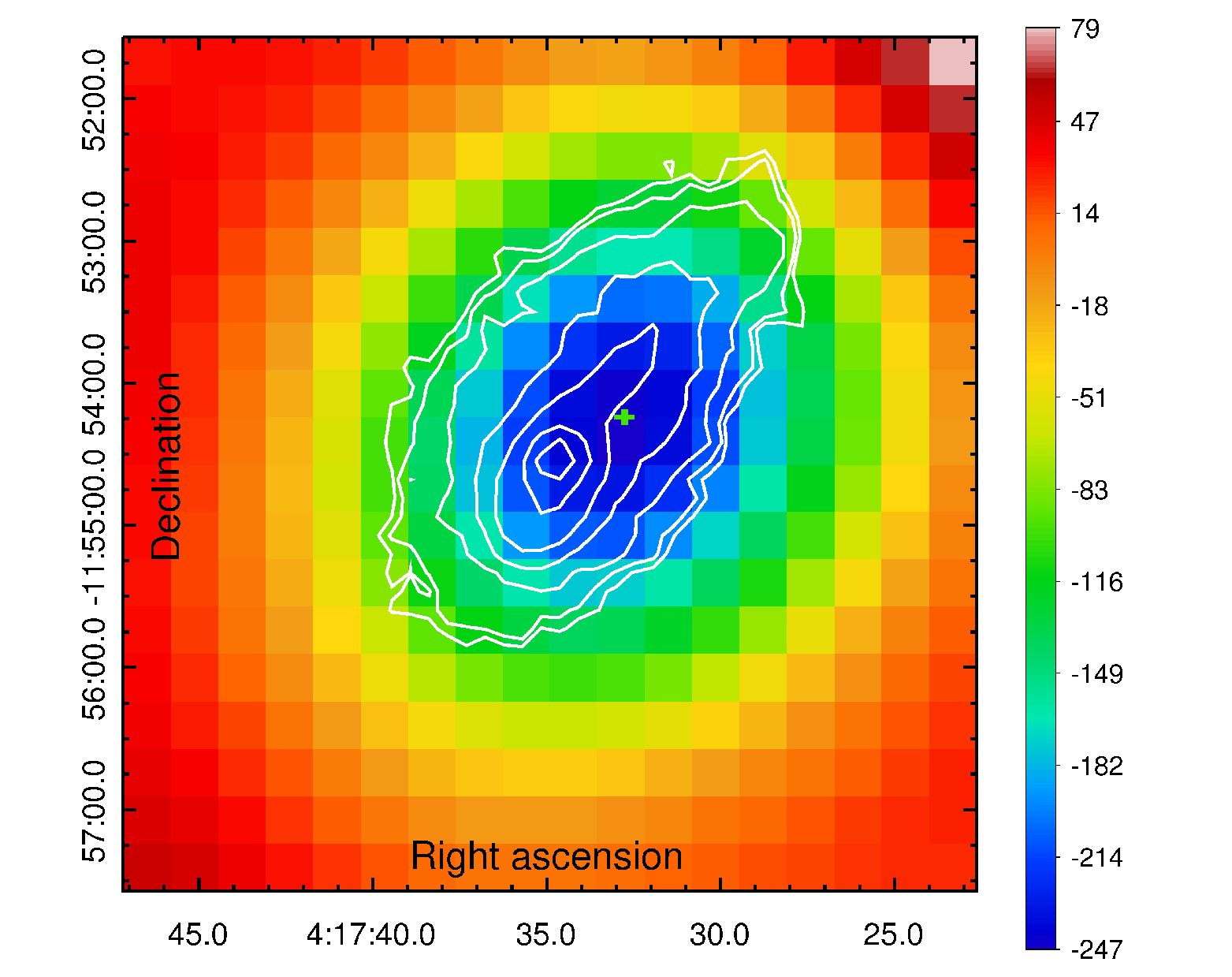}
\caption{The Sunyaev-Zel'dovich Effect (SZE) decrement, filtered map,
  at 2.1~mm wavelength (6 arcmin $\times$ 6 arcmin field) of the
  \mac~cluster obtained from the Bolocam archives. 
   The Chandra X-ray   surface brightness (white) contours have been overplotted on it for
  comparison. The green cross indicates the peak decrement position of
  the 2.1~mm SZE map.}
\label{Fig11}
\end{figure}
\subsubsection{Nature of the SE edge revealed by spectral analysis}
\label{sec:tmap}
To better understand the nature of the SE density
  discontinuity, we extract the spectra from two ``PandA''s (inner T2
  and outward T1), shown in blue in
  Fig.~\ref{fig:confirmed-jumps} ({\it left panel}). The background spectra were also extracted from similar regions from the blank sky background file \citep[see][]{2017MNRAS.472.2042P}. The extracted 
  spectra were binned (28 counts/bin), then exported to {\tt XSPEC},
  and fitted with a simple single temperature model {\tt wabs*apec},
  keeping the redshift values fixed at $z=0.44$. The Galactic $N_{H}$
  value was kept free during the fit. The best fit parameters were
  derived using $\chi^{2}$ minimisation and the errors were
  derived for the 90\% confidence level. The obtained best fit Galactic $N_{H}$ values are factor of 2.26 higher than that of reported by \citep{1990ARA&A..28..215D}.
    The best fit parameters are tabulated in Table~\ref{fit_results}.
  Considering the values of the quantities $n2/n1$
  and $T1/T2$ across the surface brightness edge, we find that the gas
  pressures across both sides of the SE edge are consistent within the
  errors. Therefore, we classify this edge as a cold front.  It turns
  out to be similar to typical sloshing cold fronts seen in many other
  systems \citep[see][for review]{2007PhR...443....1M}.  At the time
  of revising this paper, another paper on \mac\ appeared on arXiv,
  containing analysis of the {\it Chandra} observations
  \citep{2018MNRAS.476.5591B}. This paper has also reported this jump with
  $C =2.44 \pm 0.31$, which is consistent with our detection.

\subsection{Radio}
\label{subsec:Radio}

In order to further investigate the connection between the radio and
the X-ray emitting gas distributions, we overlay the GMRT 235\,MHz
radio contours on the exposure-corrected and background-subtracted
\chandra\ X-ray surface brightness image, shown in
Fig.~\ref{XOR}. This figure shows that the radio contours of \mac\ are
elongated in the SE-NW direction, closely following the X-ray
emission. The 235 MHz GMRT observation shows extended radio emission
within the SE cold front (magenta arc), as seen in some clusters
hosting a diffuse radio mini$-$halo.

\subsection{Sunyaev Zel'dovich Decrement}
\label{subsec:SZ}
Here we examine the available Sunyaev-Zel'dovich Effect (SZE)
decrement images for \mac, as the total cluster SZE signal is a good
proxy for the total mass \citep[e.g.,][]{2006ApJ...650..538N}, and is
known to trace extremely well the thermal pressure of the ICM. 
We use a publicly-available filtered SZE image from the Bolocam survey \citep{2013ApJ...768..177S}.
The final 6 arcmin $\times$ 6 arcmin, filtered Sunyaev-Zel'dovich map,
at a wavelength of 2.1~mm, is shown in Fig.\ref{Fig11}. 
 The Chandra X-ray surface brightness (white contours) have been overplotted on this
image for comparison. The green cross indicates the peak SZE decrement
position at RA $04^{h}$~$17^{m}$~$32.14^{s}$,
Dec. $-11^{o}$~$54^{'}$~$7.70^{\arcsec}$. The Chandra X-ray emission  peak is located 
at RA $04^{h}$~$17^{m}$~$34.86^{s}$, Dec. $-11^{o}$~$54^{'}$~$34.04^{\arcsec}$. 
From this comparison, it is revealed  that the strong Sunyaev-Zel'dovich decrement peak ($\Delta T \approx 500$ 
mK)  is   offset by $\sim 50^{\arcsec}$ ($\sim$ 275\,kpc)to the north-west of cluster X-ray centre in the  
direction of the radio halo extension. We  note that the  the pointing accuracy of Bolocam instrument is $3^{\arcsec}$ rms 
while for Chandra  X-ray telescope  absolute position has a error radius of  0.8 arcsec (90\% uncertainty circle), and both are 
much smaller than the significant  SZ - X-ray positional offset detectable. However,  a caveat is that  some centroid shift can 
also  be  introduced by  the  broad $\sim 60^{\arcsec}$ (FWHM) beam of the Bolocam observation.  Thus, a much finer resolution SZE
observation is  required to measure the  SZ - X-ray  offset with higher accuracy. 
The origin of this   offset can  possibly  be traced to the strong merging dynamical state of \mac, where the
  X-ray surface brightness maxima ($I_{x} \propto {\rho_{g}}^{2}
  {T_{g}}^{1/2}$) shifts to a region of the highest gas density
  ($\rho_{g}$) near the cold-front, while the SZE peak ($I_{SZ}
  \propto \rho_{g} T_{g}$) traces the region of highest thermal
  pressure of a subcluster, offset from the former due to merger  dynamics.
\subsection{Strong Lensing}
\label{subsec:Strong Lensing}
\begin{figure*}
\includegraphics[width=17cm,height=17cm]{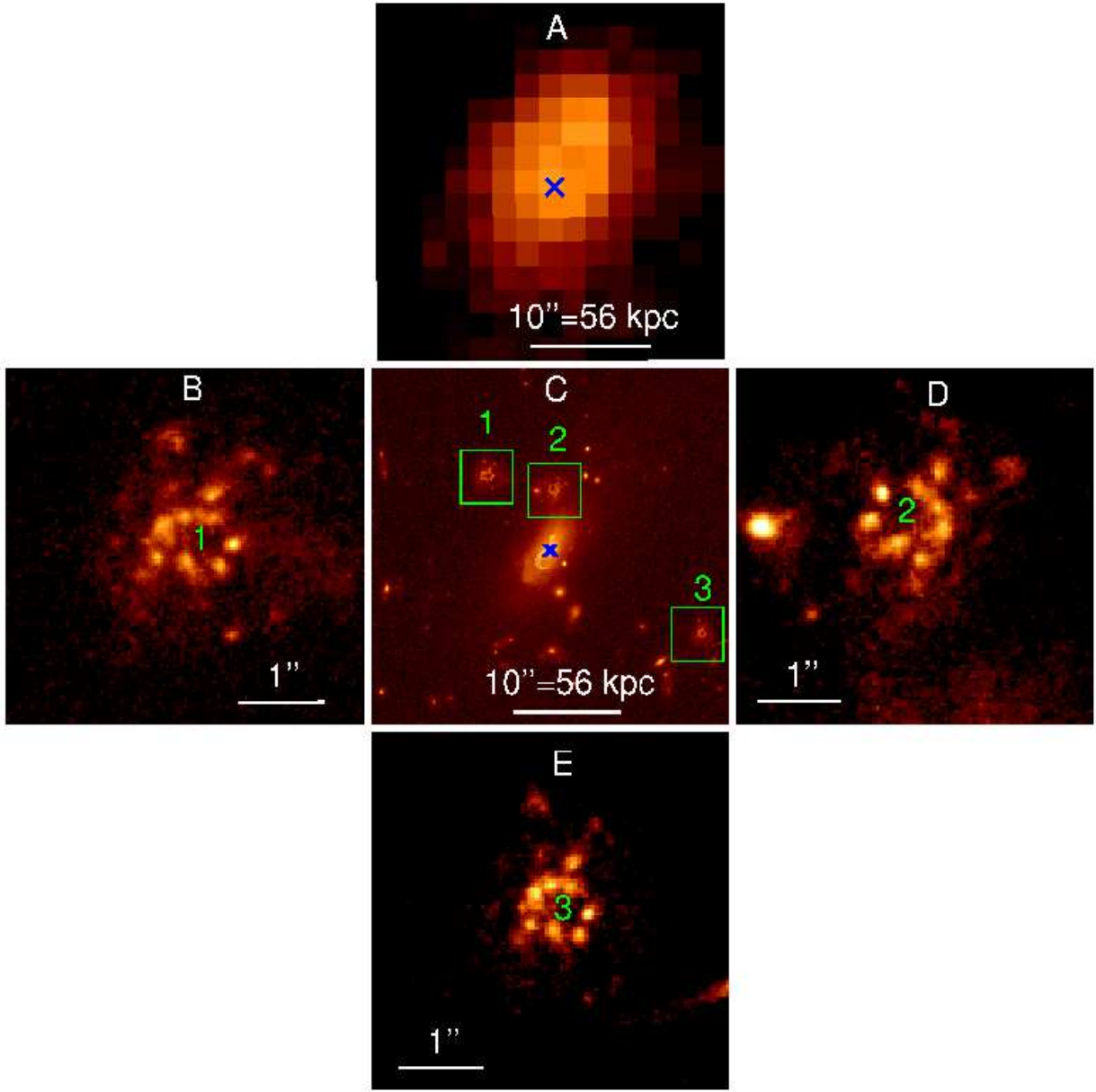}
\caption{{\it A}: The \chandra\ 0.5$-$3.0\,keV, exposure corrected, background
  subtracted \chandra\ X-ray image. The X-ray peak position is
  highlighted by a blue cross.{\it C}: HST high resolution (0.05
  arcsec/pix) (F606W) V band filter optical image of the central (1st)  BCG
  of \mac. The bright X-ray peak position is shown by blue
  cross. Peculiar circular type sources are highlighted by numbers 1
  to 3 and their zoomed images are shown in panel B, D and C,
  respectively. The FOV of panel A and C is approximately 0.5 arcmin
  $\times$ 0.5 arcmin$^{2}$~($\sim 170\times170~{\rm kpc^{2}}$), while
  panel B, D and E have angular sizes 5 arcsec $\times$ 5 arcsec ($\sim 5.5 \times
  5.5 ~{\rm kpc^{2}}$).  }
\label{Fig9}
\end{figure*}

\begin{figure*}
\includegraphics[width=8cm,height=8cm]{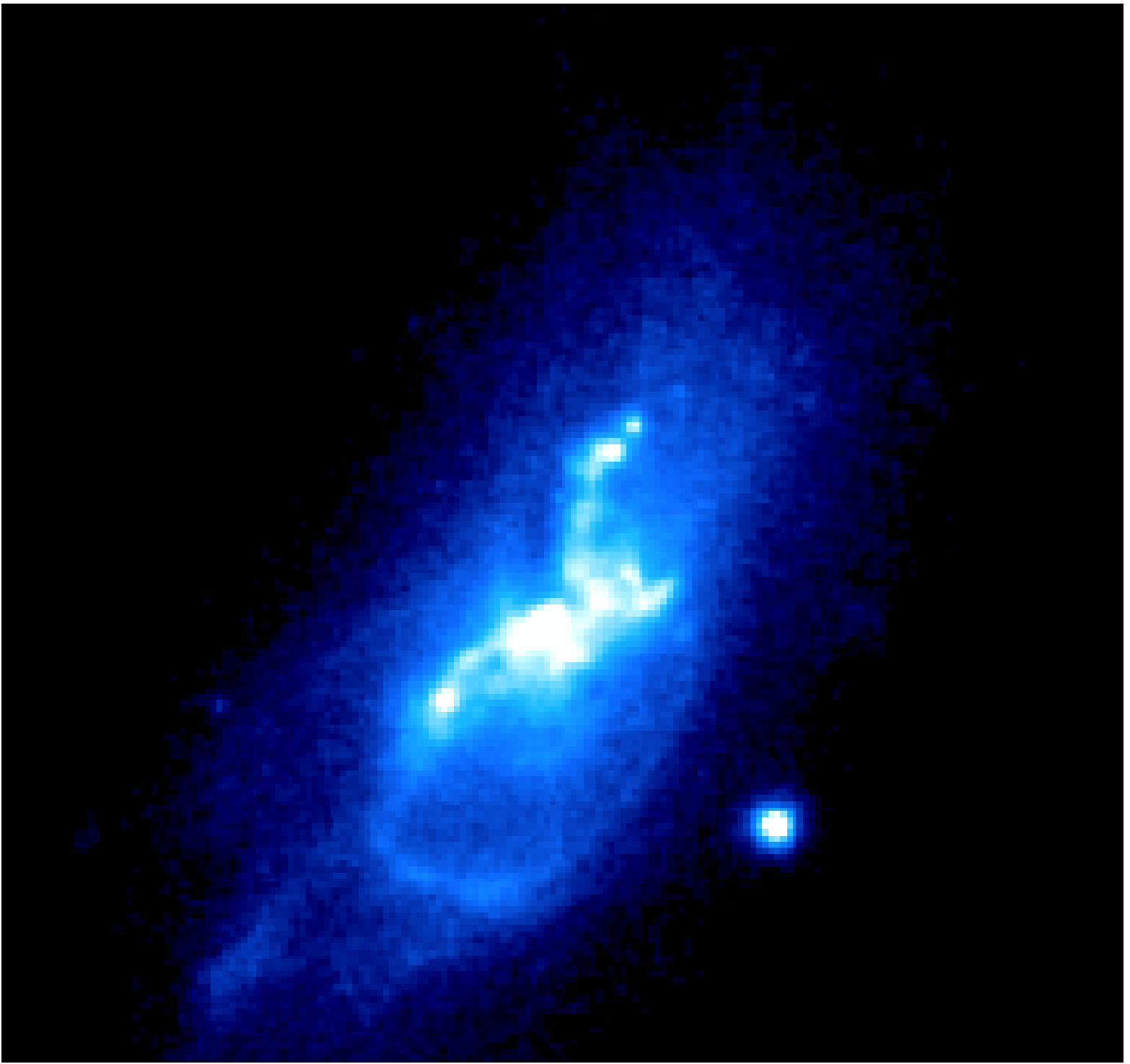}
\includegraphics[width=8.0cm,height=8.0cm]{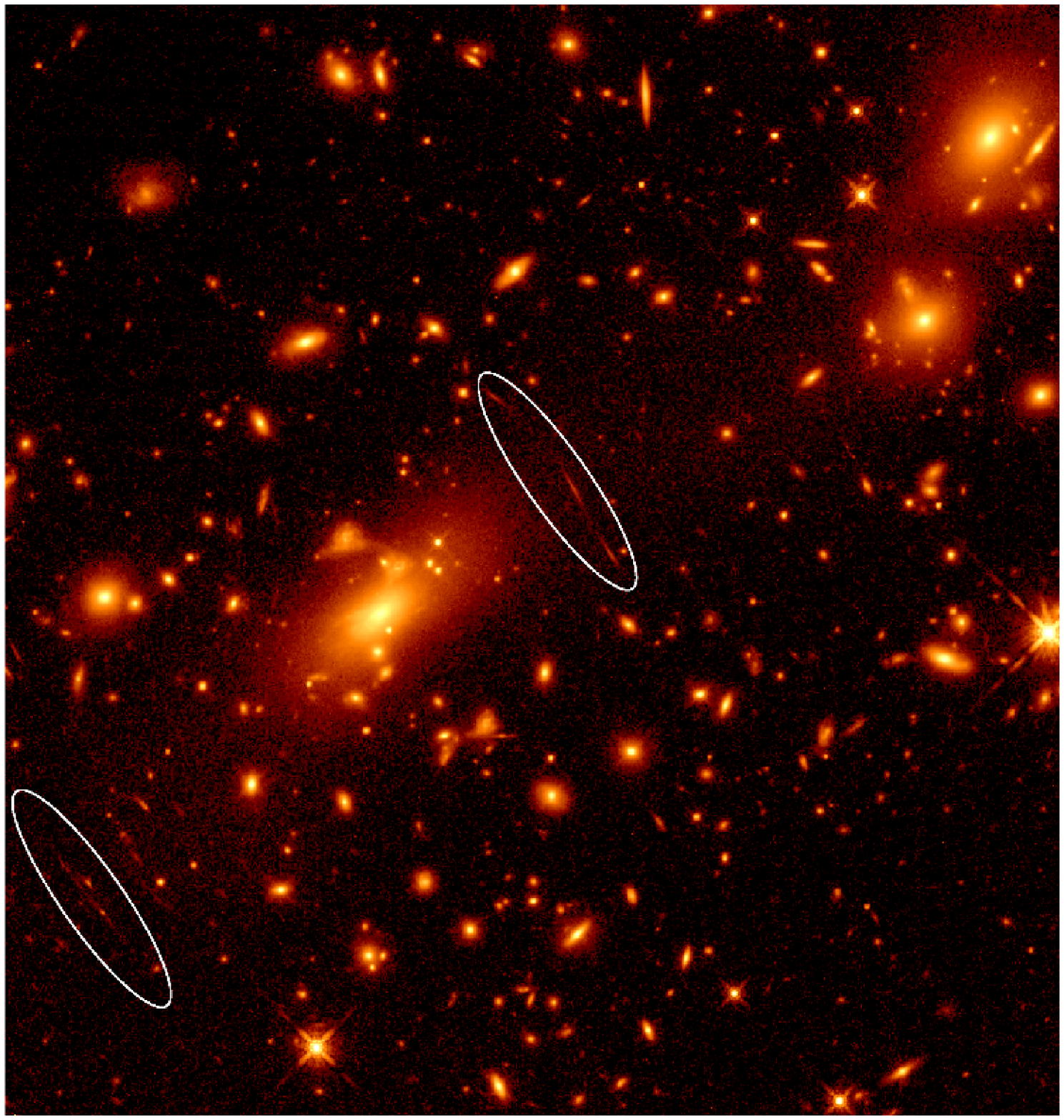}
\caption{{\it left panel}: HST 10 arcsec $\times$ 10 arcsec high$-$
  resolution UVIS V (F606W) band image. {\it Right panel}: A giant
  arc at $\sim$ 24 arcsec and a smaller arc at $\sim$ 44 arcsec are both
  evident in the HST 2.4 arcmin $\times$ 2.4 arcmin IR H band (F160W)
  filter image and highlighted by white ellipses.}
\label{Fig10}
\end{figure*}

We now investigate in more detail the $\sim$ 30 $\times$ 30
arsec$^{2}$~($\sim 170\times170~{\rm kpc^{2}}$) central region of \mac, using the \chandra\ X-ray and the high-resolution (0.05
arcsec/pix) optical images from the HST. The resultant images are shown in
Fig.~\ref{Fig9} (panels A and C). The Chandra image (panel A) shows
that the X-ray isophotes are elliptical in shape (the peak X-ray
position is highlighted by a blue ``X"). 
centre of the 1st BCG.%
 Moreover, the 1st BCG is extremely irregular
in morphology, containing several distinct filamentary
tidal-tail like features, which is very atypical of relaxed elliptical
cD galaxies found in the cores of cluster environments. These
filamentary structures are blue in colour, and possibly represent
star forming regions  related to an ongoing merger of two gas-rich
galaxies (Fig.~\ref{Fig10} left panel).  Alternatively, these complex
features might represent signatures of a vigorous active galactic
nucleus (AGN) feedback process at work.  We note that these filamentary structures
are quite similar to those found near  galaxy  NGC~1275 in the core of
the Perseus cluster \citep{1995AJ....109..960W}, or in the central galaxy of the 
cluster RXC J1504-0248 \citep{Soja18}, or near NGC~4696 in the core of the Centaurus cluster \citep{2005MNRAS.363..216C}.
This indicates that the centarl galaxy in \mac is still dynamically active and has not yet attained the relaxed 
state of a giant cD galaxy found at the cluster centers.

In addition to the complex and filamentary optical structures near the
central galaxy mentioned above, we discovered three other
interesting ring-shaped objects seen in the HST image of the
cluster. These are highlighted by green 5 arcsec $\times$ 5 arcsec
($\sim$ 5.5 $\times 5.5 ~{\rm kpc^{2}}$) boxes in Fig.~\ref{Fig9}
panel C, and marked by numbers 1 to 3. Their zoomed-in images are
shown in panels B, D and E for more clarity. Close inspection of these
objects reveals that they are strong gravitationally lensed images of
a single background galaxy, probably a distant star forming galaxy.

This interpretation is confirmed by the evident
opposite parity of images 1 and 2 (panels B and D) and the same parity
of images 1 and 3 (panels B and E) seen in Fig.~\ref{Fig9}. This means
that a critical curve in the lens plane separates images 1 and 2 and
also images 2 and 3 (images 1 and 3 are clearly of same parity).
We have not been able to locate any other possible members of this multiply-image
system.
In addition, we report two other strong lensing features: a giant arc
at 24 arcsec north of the 1st BCG and a smaller arc at 44 arcsec south of
the 1st BCG (white ellipses in Fig.~\ref{Fig10}, right panel). It is very
likely that more strong lensing features remain to be found in this
field.  Together with redshift measurements, modeling of the strong
lensing features can be used to constrain the inner mass distribution,
which is beyond the scope of the present paper.

\section{Discussion}
\label{sec:Discussion}

In spite of the relatively poor seeing and low exposure times
  of the Subaru images, we have successfully derived the mass
  distribution of the bimodal merging system MACS0417 through the weak
  gravitational lensing technique.  The recovered mass map
  (Fig.~\ref{VRI.images} ``D'') shows an elongated structure with the
  peak corresponding to the main cluster lying very close to the
  position of the first BCG. This feature is also suggested by the
  numerical density distribution of the red sequence galaxies
  (Fig.~\ref{VRI.images} ``B'' and ``C''), which also indicates that
  the dominant matter content of the subcluster surrounds the position
  of the second BCG.  Corroborating previous X-ray based estimates, we
  found a very massive system with a total mass of
  $M_{200}=13.8^{+2.6}_{-2.8} \times 10^{14}$ M$_\odot$.  Regarding
  the individual structures, the main cluster mass is evaluated as
  $M_{200}^c=11.5^{+3.0}_{-3.5} \times 10^{14}\ M_\odot$, whereas for
  the subcluster the value is $M_{200}^s=1.96^{+1.60}_{-0.95} \times
  10^{14}\ M_\odot$.
  This analysis allowed us to compute the uncertainties on the centre
  of the main cluster mass distribution, and to compare that quantity
  with both the ICM (traced by its X-ray emission) and the
  distribution of the constituent galaxies (with the first BCG as
  their centroid). The identification of a possible detachment among
  these quantities is the base for models which explore a possible
  self-interaction behaviour of the dark matter
  \citep[e.g.][]{2015Sci...347.1462H}. However, our results have shown
  that the position of the main cluster mass centre, the corresponding
  X-ray peak and the 1st BCG are all comparable within 99\% c.l.
  (Fig.~\ref{VRI.images} ``E''), meaning that this structure has
  retained its gas content after the pericentric passage of the
  subcluster.

 As far as the subcluster is concerned, a related ICM clump is
   not detected, suggesting that it can be disrupted due to the
   interaction with the main cluster. Therefore, \mac~can be
   classified as single dissociative merging cluster \citep{dawson},
   where only the most massive component has retained its gas content
   as seen, e.g., in A1758 \citep[][]{Monteiro-Oliveira17a}.  Future
   combined strong plus weak lensing analysis will improve our
   understanding of the mass distribution in the inner part of
   MACS0417.  Additionally, future radial velocity surveys of the
   cluster members will also enable us to fully address the current
   dynamical state of this merger and determine the position of the
   merger axis in relation to the plane of the sky. In addition, the
   reconstructed dynamical state can be used as an input for
   tailor-made hydrodynamical simulations aiming to explain the
   observed features in this interesting target.

The X-ray surface brightness profiles and detected edges imply that
the density discontinuities seen in the ICM/IGM are related to either
shock fronts or cold fronts, depending on the direction of the temperature
jump \citep[e.g.,][]{2007PhR...443....1M}.  Now, it is believed that
the radio halos result from the re-acceleration of mildly relativistic
electrons during a merger event between galaxy clusters
\citep[e.g.,][]{2001MNRAS.320..365B,2001ApJ...557..560P}. \mac~hosts a radio halo elongated in the
southeast $-$ northwest (hereafter SE$-$NW) direction
\citep{2011JApA...32..529D,2017MNRAS.464.2752P}.

We searched for surface brightness edges near the location of the
radio halo. A sharp surface brightness edge is clearly visible in
Fig.~\ref{fig1}, located 45 arcsec $\sim$ (255\,kpc) towards SE
direction from the center of the cluster, where a compact radio core
of an AGN is also visble, which is superposed on the above radio halo.
This unresolved radio core is possibly associated with the brightest
central knot (an active galaxy nucleus) of the 1st BCG visible on the
HST image, measured at position RA $04^{h}$~$17^{m}$~$34.7^{s}$,
Dec. $-11^{o}$~$54^{'}$~$32.3^{\arcsec}$.  The compact X-ray core is
also visible in {\it Chandra} X-ray image, measured at position RA
$04^{h}$~$17^{m}$~$34.56^{s}$,
Dec. $-11^{o}$~$54^{'}$~$30.0^{\arcsec}$. The SE edge is best fitted
by a broken power-law model with a density compression factor
$C=2.21\pm0.12$, and there is a temperature jump evidence that it is associated
with a cold front. \mac~is a compact cool core merging
cluster. Therefore, the cold front is most likely a sloshing driven
cold front, similar to that of Abell 2142 \citep{2013A&A...556A..44R}
and A2052 \citep{Machado15}.  From Fig.~\ref{fig1}, we can see that
excess X-ray emission toward the SE direction is formed by the ICM of
the core sloshing around the centre of \mac. We expect this excess
emission region to be metal-rich and cooler than that of nearby
regions from the centre of the cluster.  Analyses of the temperature
and metal abundance reveal that the region corresponding to the excess
X-ray emission, associated to the SE cold front presented in
\S.~\ref{sec:tmap}, indeed has a lower temperature and higher metal
abundance, $Z=0.54\pm0.14$ \Zsun, than other regions located at same
distance from the centre, $Z=0.29\pm0.12$ \Zsun.

\citet{2017MNRAS.464.2752P} found that the spectral
  index\footnote{The radio flux density, $S$, is proportional to
    $\nu^\alpha$, where $\nu$ is the frequency of the observation and
    $\alpha$ is the spectral index.} of the diffuse radio emission
  between 610 and 1575~MHz is $\alpha\sim -1.72$, and classified this
  source as an ultra-steep spectrum giant ($\sim$ 1.1\,Mpc) radio
  halo.  However, considering the central temperature, entropy and
  cooling time \citep {2009ApJS..182...12C,2017ApJ...841...71G}, we have
  classified this source as a cool-core giant radio-halo-host cluster
  similar to CL1821+643 \citep{2014MNRAS.444L..44B}, A2390 and A2261
  \citep{2017MNRAS.466..996S}.  This makes \mac\ one of the
  exceptionally rare cool core clusters hosting an ultra steep spectrum
  radio halo. Past studies have shown that giant radio halos are
  usually found in massive mergers and not in cool core clusters. On
  the other hand, a different class of radio sources, namely `radio
  mini haloes' are common in cool core clusters
  \citep{2017ApJ...841...71G}. This brings up the possibility that the
  radio emission in \mac\ could be generated by the same physical
  mechanism responsible for `radio mini haloes', but on a much larger
  physical scale of 1.1 Mpc.

A spatial correlation has been observed between cold
  fronts and radio minihalos detected in a few galaxy clusters
  \citep{2008ApJ...675L...9M}, suggesting a connection between the
  sloshing motion of the gas and the origin of this kind of radio
  sources \cite{2013ApJ...762...78Z}. With the use of high-resolution
  MHD simulations, it has been shown that gas sloshing can cause
  turbulent re-acceleration of relativistic electron seeds (e.g., from
  past AGN activity) and can produce diffuse steep-spectrum radio
  emission within the region of the sloshing cold front. The SE cold
  front and the radio halo in \mac\ next to the sloshing cold front,
  extending towards the NE direction behind it, is quite consistent
  with this established scenario.
  
  Another possibility is that the \mac\ system could be
  an intermediate class, where a radio minihalo has been switched off
  because of the turbulent motions generated during the merger event,
  and the relativistic electrons have been transported across larger
  distances than usual \citep{2014IJMPD..2330007B}. In this case, one
  should assume that the merger has dissipated enough energy to
  re-accelerate relativistic electron seeds (e.g., from past AGN
  activity), but not enough to disrupt the cool core.

\section{Conclusion}
\label{sec:conclusion}

We present observations of the merging cluster system \mac\ using
archival data obtained with \chandra\ (X-ray), Subaru, HST (optical),
GMRT (low-frequency radio) and Bolocam (SZE 2.1~mm). From our analysis presented in
this paper, we can draw the following conclusions.

\begin{enumerate}

\item The weak lensing analysis of \mac, which is the first in
  the literature considering the system as the combination of a main
  cluster plus a subcluster, yields a mass ratio of $\sim$
  6:1 of the two dark matter haloes.
  The individual masses are $M_{200}^c=11.5^{+3.0}_{-3.5} \times
  10^{14}\ M_\odot$ (main cluster) and $M_{200}^s=1.96^{+1.60}_{-0.95}
  \times 10^{14}\ M_\odot$ (subcluster) leading to a total cluster
  mass of $M_{200}=13.8^{+2.6}_{-2.8} \times 10^{14}\ M_\odot$.\\

\item \mac~is classified as a dissociative merger since the
  main cluster has retained its gaseous content as shown by the
  detected spatial coincidence (within 99\% c.l.) of the distribution
  of the dark matter, ICM and member galaxy components. On the other
  hand, the subcluster seems to have had its gas content disrupted due
  to the merger.\\

\item The optical image of the first BCG shows interesting filamentary
  substructure, which is atypical of normal BCGs and may indicate an
  ongoing merger between the two galaxies.\\

\item We confirm a surface brightness edge to the east-south direction
  at a distance of 45 arcsec ($\sim$ 255\,kpc) from the centre of
  \mac, also reported by \cite{2018MNRAS.476.5591B}.\\

\item The spectral analysis across the inner and outer regions of the
  edge has allowed us to confirm that this sharp emission edge is a
  cold front.  The overall structure, surface brightness profile,
  temperature and metal abundance values point towards the presence of
  a sloshing induced cold front in this system. \\



\item We detect three components of a possible multiply-imaged
  strongly lensed background star forming galaxy, plus two previously
  unreported arcs which are all indications of strong lensing in this
  system. Strong lensing analysis of \mac\ will lead to a clearer understanding
  of this cluster.

\item We find that the peak of the Sunyaev-Zel'dovich decrement of
  \mac\ is displaced from its X-ray peak, which we interpret to be a
  consequence of the merger dynamics in the system.

\end{enumerate}

\section*{Acknowledgments}
MBP gratefully acknowledges the support from following funding
schemes: Department of Science and Technology (DST), New Delhi under
the SERB Young Scientist Scheme (SERB/YSS/2015/000534), Department of
Science and Technology (DST), New Delhi, and the INSPIRE faculty
Scheme (DST/INSPIRE/04/2015/000108).  MBP also wishes to acknowledge
with thanks the support received from IUCAA, Pune, India in the form
of their visiting Associate programme.  RMO thanks very much the
faculty members and staff of the Astronomy Department/Institut of
Physics of the Federal University of Rio Grande do Sul for their
lovely hospitality and the provided support which made possible the
continuity of this research. RMO also thanks Prof. Eduardo S. Cypriano
(IAG/USP) for share his computational resources and Prof. Cl\'audia
L\'ucia M. Oliveira (IAG/USP) for the administrative support.  ML
acknowledges CNRS for its support. ML acknowledges CNRS and CNES for support.
 This research has made use of the
data from {\it Chandra} Archive. Part of the reported results are
based on observations made with the NASA/ESA Hubble Space Telescope,
obtained from the Data Archive at the Space Telescope Science
Institute, which is operated by the Association of Universities for
Research in Astronomy, Inc.,under NASA contract NAS 5-26555. This
research has made use of software provided by the Chandra X-ray Centre
(CXC) in the application packages CIAO, ChIPS, and Sherpa. This work
is based on observations taken by the RELICS Treasury Program (GO
14096) with the NASA/ESA HST, which is operated by the Association of
Universities for Research in Astronomy, Inc., under NASA contract
NAS5-26555. This research has made use of NASA's Astrophysics Data
System, and of the NASA/IPAC Extragalactic Database (NED) which is
operated by the Jet Propulsion Laboratory, California Institute of
Technology, under contract with the National Aeronautics and Space
Administration.  Facilities: Chandra (ACIS), HST (ACS). This work also
is based on data collected at Subaru Telescope, which is operated by
the National Astronomical Observatory of Japan.
\def\aj{AJ}%
\def\actaa{Acta Astron.}%
\def\araa{ARA\&A}%
\def\apj{ApJ}%
\def\apjl{ApJ}%
\def\apjs{ApJS}%
\def\ao{Appl.~Opt.}%
\def\apss{Ap\&SS}
\def\aap{A\&A}%
\def\aapr{A\&A~Rev.}%
\def\aaps{A\&AS}%
\def\azh{AZh}%
\def\baas{BAAS}%
\def\bac{Bull. astr. Inst. Czechosl.}%
\def\caa{Chinese Astron. Astrophys.}%
\def\cjaa{Chinese J. Astron. Astrophys.}%
\def\icarus{Icarus}%
\def\jcap{J. Cosmology Astropart. Phys.}%
\def\jrasc{JRASC}%
\def\mnras{MNRAS}%
\def\memras{MmRAS}%
\def\na{New A}%
\def\nar{New A Rev.}%
\def\pasa{PASA}%
\def\pra{Phys.~Rev.~A}%
\def\prb{Phys.~Rev.~B}%
\def\prc{Phys.~Rev.~C}%
\def\prd{Phys.~Rev.~D}%
\def\pre{Phys.~Rev.~E}%
\def\prl{Phys.~Rev.~Lett.}%
\def\pasp{PASP}%
\def\pasj{PASJ}%
\def\qjras{QJRAS}%
\def\rmxaa{Rev. Mexicana Astron. Astrofis.}%
\def\skytel{S\&T}%
\def\solphys{Sol.~Phys.}%
\def\sovast{Soviet~Ast.}%
\def\ssr{Space~Sci.~Rev.}%
\def\zap{ZAp}%
\def\nat{Nature}%
\def\iaucirc{IAU~Circ.}%
\def\aplett{Astrophys.~Lett.}%
\def\apspr{Astrophys.~Space~Phys.~Res.}%
\def\bain{Bull.~Astron.~Inst.~Netherlands}%
\def\fcp{Fund.~Cosmic~Phys.}
\def\gca{Geochim.~Cosmochim.~Acta}%
\def\grl{Geophys.~Res.~Lett.}%
\def\jcp{J.~Chem.~Phys.}%
\def\jgr{J.~Geophys.~Res.}%
\def\jqsrt{J.~Quant.~Spec.~Radiat.~Transf.}%
\def\memsai{Mem.~Soc.~Astron.~Italiana}%
\def\nphysa{Nucl.~Phys.~A}%
\def\physrep{Phys.~Rep.}%
\def\physscr{Phys.~Scr}%
\def\planss{Planet.~Space~Sci.}%
\def\procspie{Proc.~SPIE}%
\bibliographystyle{mn.bst}
\bibliography{mybib.bib}
\end{document}